\documentclass[sigconf]{acmart}

\usepackage[utf8]{inputenc}

\AtBeginDocument{%
  }
\copyrightyear{2026}
\acmYear{2026}
\setcopyright{cc}
\setcctype{by}
\acmConference[ASSETS '26]{The 28th International ACM SIGACCESS Conference on Computers and Accessibility}{October 25--28, 2026}{Vila Nova de Gaia, Portugal}
\acmBooktitle{The 28th International ACM SIGACCESS Conference on Computers and Accessibility (ASSETS '26), October 25--28, 2026, Vila Nova de Gaia, Portugal}
\acmDOI{10.1145/3797867.3829044}
\acmISBN{979-8-4007-2521-0/2026/10}
\begin{document}

\makeatletter
\def\@ACM@copyright@check@cc{}
\makeatother   
\title{Me, Myself, and My Voice: Exploring Cultural and Linguistic Identity in AAC AI-generated Voices}

\author{Tobias Weinberg}
\affiliation{%
\department{Computer Science}
  \institution{Cornell Tech}
  \city{New York}
  \country{USA}}
\email{tmw88@cornell.edu}

\author{Aaleyah Lewis}
\affiliation{%
\department{Paul G. Allen School of Computer
Science and Engineering}
  \institution{University of Washington}
  \city{Seattle, Washington}
  \country{USA}}
\email{alewis9@cs.washington.edu}

\author{Ricardo E. Gonzalez Penuela}
\affiliation{%
\department{Information Science}
  \institution{Cornell Tech}
  \city{New York}
  \country{USA}}
\email{reg258@cornell.edu}

\author{Weicong Hong}
\affiliation{
\institution{Cornell Tech}
\city{New York}
\country{USA}
}
\email{wh528@cornell.edu}

\author{Jennifer Mankoff}
\authornote{Both authors contributed equally to the paper.}
\affiliation{%
\department{Allen School of Computer Science
and Engineering}
  \institution{University of Washington}
  \city{Seattle, Washington}
  \country{USA}}
\email{jmankoff@acm.org}

\author{Thijs Roumen}
\authornotemark[1]
\affiliation{%
\department{Information Science}
  \institution{Cornell Tech}
  \city{New York}
  \country{USA}}
\email{thijs.roumen@cornell.edu}
\orcid{0000-0003-2042-6597}

\renewcommand{\shortauthors}{Weinberg et al.}

\begin{abstract}
Voice is a central element of identity. We recognize people by their voice, and we uniquely express who we are with it. For people who rely on augmentative and alternative communication~(AAC) systems, such as speech-generating devices~(SGD), the device's voice becomes an identity marker others associate with them. Yet, it is hard to find a voice that truly aligns with one's identity both linguistically and culturally. 
Although modern AI-generated voices can reproduce diverse accents and speaking styles, AAC users still lack accessible ways to articulate how they want an identity-aligned voice to sound like. We first conducted a survey of AAC users (across eight countries) to characterize current voice representation, finding that non-binary, transgender, and non-US-born respondents rated their current voice support identity alignment consistently lower than other respondents. To examine how AAC users respond to voices designed to reflect their cultural identity, we built a tool that elicits cultural markers through guided questions and generates personalized voice candidates for participants to hear and reflect on. After participants heard the voices, we interviewed them to examine what it means for a voice to feel culturally representative, how they interpreted voices with cultural connotations, and how these voices shaped their sense of identity and agency. 
Our findings show that cultural voice alignment runs deeper than accent or language alone; it touches on belonging, self-recognition, and what it means to be heard as who you are.

\end{abstract}



\begin{teaserfigure}
  \centering
  \includegraphics[width=0.55\textwidth]{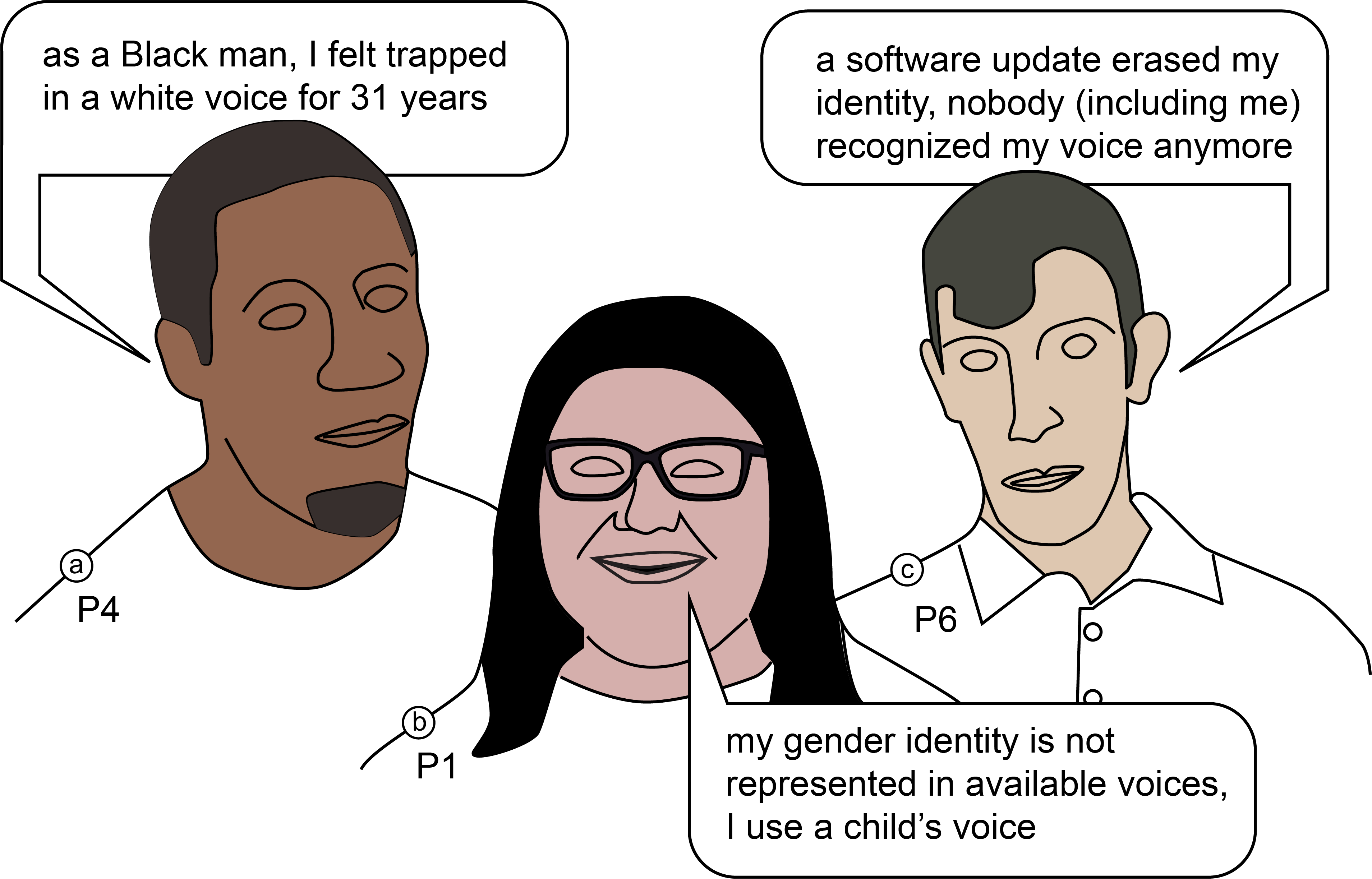}
  \caption{We explore cultural identity in AAC voice generation. Participants in our study reported several forms of misalignment, misrepresentation, and dependence on their synthetic voices. Creating voices that align culturally and emotionally with their users is crucial to reinforce a sense of self and identity for AAC users.}
  \Description{An illustration of three AAC users with speech bubbles showing quotes from the study. On the left, a Black man labeled P4 says 'as a Black man, I felt trapped in a white voice for 31 years.' In the center, a white person with glasses labeled P1 is smiling. P1's speech bubble reads 'my gender identity is not represented in available voices, I use a child's voice.' On the right, a white man labeled P6 says 'a software update erased my identity, nobody (including me) recognized my voice anymore.' The image introduces the paper's themes of cultural misalignment and identity in AAC voice use.}
  \label{fig:teaser}
\end{teaserfigure}


\maketitle

h\section{Introduction}

How would you describe your voice? What makes it unique? Voice plays a central role in how people present themselves socially: it carries cues about culture\footnote{In this paper, we use the term ``culture'' to refer to regional heritage, values, and traditions that shape communication styles in the communities people are part of, but we acknowledge that culture encompasses many additional dimensions (e.g., religion). }, locale, personality, and belonging; it shapes how others perceive speakers, and how they perceive themselves. For people with speech impairments, who rely on augmentative and alternative communication~(AAC) technology with synthetic voices,  the voice of the device becomes their public voice: the voice their colleagues, friends, and strangers associate with them~\cite{wickenden_whose_2011, judge2013perceptions}. AAC users are a heterogeneous population; complex communication needs arise from congenital and early-onset conditions (e.g., cerebral palsy, autism), degenerative diseases (e.g., ALS), and acquired impairments (e.g., stroke, traumatic brain injury).

\textit{Voice}, in the context of this paper, refers to the acoustic properties of speech, accent, timbre, prosody, and speaking rate, and how words are spoken (i.e., prosodic voice). This ``prosodic voice'' is closely tied to how individuals express identity and maintain a sense of self, which is why the loss of one’s native voice can disrupt both self-concept and social participation~\cite{nathanson2017native}. At the same time, the social context surrounding AAC use, including communication partner attitudes and perceptions of assistive technologies, plays an important role in determining how these systems are adopted and used in everyday life~\cite{baxter2012barriers}. Research on speech and social interaction shows that vocal characteristics function as social signals that influence how speakers align with or differentiate themselves from others in conversation~\cite{pardo2012reflections}. These dynamics extend to synthetic voices as well: studies examining stakeholder perceptions of synthesized voices across multiple languages show that voice characteristics affect whether they are perceived as natural, appropriate, and socially acceptable within different communities ~\cite{terblanche2025you}. 

For AAC users, vocal identity has historically been reduced to a small set of synthetic voices; these voices fail to reflect the linguistic and cultural diversity of their users, creating a mismatch between the user’s identity and the voice that represents them in conversation~\cite{judge2013perceptions}. Advances in synthetic speech have made it technically feasible to generate voices that reflect regional accents and other sociolinguistic variation~\cite{michel2025bias, terblanche2025you}. These voices may be specified using voice cloning, when training data is available, or via textual description.  
However, cloning is not an accessible option for many AAC users, and current synthesis systems often lack the representational range to produce the voice a user wants, with American and British accents dominating \cite{michel2025bias}. Of particular relevance for our target population, people who do not use voice cloning, the multi-dimensional nature of voice (accent, timbre, prosody, and tempo) is difficult to describe in words.
This raises a set of questions we set out to study: How well do current synthetic voices represent the cultural and linguistic\footnote{We refer to linguistic identity as how individuals and groups express, negotiate, and signal identity through language use, including language, dialect, and broader speaking styles \cite{LinguisticIdentity}.} identities of AAC users? How do AAC users describe the voice they want when that voice is tied up with cultural identity, belonging, and self-presentation? And what does it mean to be heard as yourself?

To answer these questions, we conducted a mixed-method study. We examine how AAC users imagine, articulate, and evaluate culturally aligned synthetic voices. We started with a global survey with 53 AAC users to understand how well current synthetic voices represent their cultural and linguistic identities and what barriers users face in personalizing them. We found that while most participants rated their current voice as functionally adequate, ease of personalization was rated negatively by a majority of respondents. Building on these findings and the lived experience of the first author as a long-term AAC user who communicates in different languages, we conducted qualitative interviews with 6 AAC users to explore what it means for users to have a voice that feels culturally (mis)aligned. We show that cultural voice alignment extends beyond accent or language alone and includes self-recognition, belonging, and the feeling of being heard as who you are. 

\section{Contributions, Benefits, and Limitations}

This paper makes the following contributions:

\begin{itemize}
    \item \textbf{An international survey of cultural voice representation in AAC.} Through a survey of 53 AAC users, we report on how well current synthetic voices represent users' cultural and linguistic identities, and the barriers that make voice personalization difficult in practice.
    \item \textbf{Empirical findings on how AAC users articulate and interpret culturally aligned synthetic voices.} Through a guided voice-elicitation workflow and follow-up interviews, we show how participants described the voices they wanted and how they reacted to the voices they heard, in relation to authenticity, familiarity, belonging, self-recognition, and social acceptability.
    \item \textbf{Design implications for future voice-generation.} We derive implications for systems that support culturally grounded and identity-affirming voice creation, including the need for reflective elicitation, iterative refinement, and user control over how cultural cues are represented in synthetic voices.
\end{itemize}

This work contributes early evidence that voice personalization in AAC should be understood as an identity-centered design problem, not only a technical synthesis problem. Our findings help surface what AAC users want culturally aligned voices to represent, why those preferences are often difficult to specify directly, and how generated voices can support or disrupt a sense of self. 

\textbf{Limitations.} The number of participants in our study is small; furthermore, we acknowledge that there is a broad range of AAC users. We note that the experiences described in this paper are specific to adult AAC users without intellectual disabilities who use speech-generating devices with synthetic voices; users who have never reliably used oral speech will relate differently to questions of voice identity than those who acquired AAC after a period of natural speech. In the context of this paper, we narrow this down to AAC users who use speech-generating devices~(SGDs), and we do not claim generalizability across cultures or AAC users.  Our findings reflect empirical observations as anecdotal evidence. While our survey reaches a broader audience, we still highlight that culture and representation are highly personal and thus refrain from drawing aggregate conclusions. 
Our diverse participants included individuals with multiple marginalized identities (e.g., Hispanic, black, non-binary, and multilingual). We acknowledge that a survey limits our ability to capture the nuance of these intersecting experiences. Our survey did not collect information about participants' underlying diagnoses or disability types, which limits our ability to fully characterize the range of AAC users represented in our sample. We conducted quantitative analysis along single axes (e.g., non-binary vs. cisgender; U.S. vs. non-U.S.), which may obscure how multiple identities jointly shape AAC experiences and perceptions of voice representation. 
Finally, this study does not assess long-term adoption; future work should explore longitudinal use, broader populations, and more iterative forms of voice co-creation over time.
\section{Positionality Statement}

The lead author has been an AAC user for over 13 years; he lost his ability to speak at fifteen. Growing up in Argentina, he had no guidance on what devices existed or how to use them. When he first began using a speech-generating device, the available Spanish voices reflected neither how he spoke nor where he was from: the options were Mexican or Spain Spanish, and none of them sounded like anyone he knew. \textit{“I didn't like sounding different from all my friends. Internally, I felt like an outsider.”} This research is motivated by the lived experience of being handed a voice that was technically the right language but culturally the wrong one.

We disclose this positionality because it shapes how the work was conducted and how the findings are interpreted. The lead author's firsthand experience with cultural voice misalignment informed the design of the survey, the questions asked in interviews, and the sensitivity with which participant responses were read. At the same time, we acknowledge that his experience as an Argentine Spanish speaker is one specific form of misalignment among many, and that participants in this study brought cultural identities and voice expectations that differ substantially from his own.

The lead author's experience extends beyond the design of this study. Over the past year, he has used an AI-generated voice designed to reflect his own cultural and linguistic identity as his primary AAC voice for presentations and public speaking. This experience has been one of ongoing iteration: prompt refining, listening, adjusting, and listening again, in search of a voice that sounds like who he is. The first time he heard a voice that reflected his Argentine background, after more than ten years of speaking through voices that sounded nothing like him, he felt a happiness he had not expected. For the first time since losing his native voice, he felt represented by the sound coming out of his device. That experience gave him a new confidence and a renewed pride in his roots that he had not realized were missing.

The experience has also revealed the fragility of that alignment. When \textit{ElevenLabs} \footnote{ElevenLabs is a commercial AI voice synthesis platform that generates speech from text-based descriptions.} updated their underlying model, the voice he had carefully refined changed, sometimes shifting accent mid-sentence, sometimes varying unpredictably across regenerations. What had felt like his voice became unreliable. This instability points to something important: cultural voice alignment in AAC is not a one-time configuration but an ongoing relationship between a user, a voice, and a technology that can change without warning. 
\section{Related work}

``Voice'' refers to two related but distinct phenomena: lexical and prosodic voice. Drawing on foundational work in phonetics, including Laver \cite{laver1994principles}, we use voice to refer to the acoustic properties of speech and how words are spoken (i.e., prosodic). When referring to the content or style of what is said to express oneself (e.g., wording, syntax), we explicitly specify as lexical voice.
 
These dimensions are entangled in AAC, where generative AI may contribute both simultaneously. 

Voice is a strong carrier to express identity, and is deeply tied to autonomy and social perception.  
\citet{nathanson2017native} argues that voice loss constrains expression of the self, and that the ethical principles of autonomy establish a moral responsibility to invest in personalized voice technology. For AAC users, the device's voice becomes the public voice others associate with them, shaping how users are perceived during interaction~\cite{judge2013perceptions}. \citet{baxter2012barriers} identify the voice and lexical properties of the device as key factors in AAC adoption, noting that available options rarely meet users' needs. \citet{wickenden_whose_2011} shows how teenage AAC users' voices emerge in co-constructed conversations, which is central to how they see themselves and are seen by others, and that having a different voice carries implications for identity and social recognition. Beyond AAC, ~\citet{pardo2012reflections} argues that speakers' convergence or divergence in how they sound depends on the social context of the interaction, highlighting that voice is an active social signal rather than a neutral carrier of content. Our work builds on these insights by examining not just whether a voice supports communication, but whether it sounds like who the user is.

 To understand what it means for a voice to sound like who a user is, three bodies of work are relevant: (1)~Self-expression and Agency in AAC Use, (2)~Voice Personalization Technology, and (3)~Identity Representation in AAC Systems.

\subsection{Self-expression and Agency in AAC Use}

\textit{Conversational agency}, an individual's capacity to express and achieve their goals in conversation, is central to self-expression: without the capacity to say what one wants, when one wants, in the way one wants, identity in interaction is constrained before it even begins. Conversational agency is not solely a function of speed or clarity, but emerges through interaction with the context of the conversation~\cite{valencia_conversational_2020, valencia2024compa}. It is shaped by factors that engage both lexical and prosodic voice, including social constraints (such as the norms and expectations of a given interaction), the pace of conversation~\cite{kane_at_2017}, and the responsiveness of the used AAC technology, its vocabulary, and the connection to the context of interaction~\cite{curtis2022state, curtis2024breaking}. For example, a user in a fast-paced group conversation faces different agency constraints than one in a quiet one-on-one exchange: the social setting, the communication partner's familiarity, and the time pressure each shape what can be expressed and how~\cite{, higginbotham2013slipping, higginbotham2016time}.

Identity and agency are deeply linked: the words one is able to produce shape how one is recognized by others. \citet{kane_at_2017} have shown the complex dynamics that AAC users navigate to express themselves and retain their identity. This dependency raises concerns about control of expression and authorship: AAC users and practitioners worry that AI systems can homogenize AAC outputs and blur the line between the user's own voice and system-generated output~\cite{griffiths2024use, griffiths2025ai, weinberg2025robot}. 
Prior work outside of AAC literature suggests that these concerns are not merely speculative. \citet{abdulhai2026llms} demonstrate this risk, showing that AI systems used for writing assistance do more than alter lexical voice: beyond stylistic flattening, systems can systematically shift the intended meaning of users’ text, steering revisions toward normative semantic directions, rather than preserving individual expression.

For AAC users, these identity concerns are further complicated by a practical tradeoff: in real-time conversation, full agency over one's lexical voice often conflicts with conversational timing \cite{ valencia_less_2023}. \citet{weinberg2025why} showed that in time-sensitive contexts, AAC users accept reduced agency in favor of delivering humorous comments faster. They found that humorous comments functioned as a form of backchanneling, communicating presence and turn-taking in the conversation. Deeper investigation into backchanneling practice of AAC users demonstrated that these unique cues form part of a broader AAC micro-culture, supporting self-expression~\cite{weinberg2025one}. Together, this suggests that conversational agency in AAC is the ability to participate on time, retain authorship, and remain recognizable as oneself under the constraints of real interaction.
\subsection{Voice Personalization Technology}

Voice personalization in AAC has primarily been approached through voice banking, recording a person's speech to build a synthetic voice personalized to an individual's prosodic voice~\cite{yamagishi2012speech, cave2021voice, veaux2013towards, veaux2011voice}. However, voice banking presupposes access to usable natural speech recordings, and is therefore not an option for many AAC users, including those who have never reliably used oral speech. \citet{mills2014towards} describe the \textit{VocaliD} approach, which addresses a key limitation of traditional voice banking: for users with profoundly limited speech motor control, who cannot produce clearly articulated recordings, \textit{VocaliD} extracts prosodic properties from the target talker's residual vocalizations and applies them to a surrogate donor's database, generating a voice that carries the vocal identity of the target with the clarity of the surrogate. Crucially, their work also highlights that Text To Speech~(TTS) options on AAC devices are limited and that several individuals in a group often use the same synthetic voice, a lack of customization that may limit technology adoption and social integration. However, scholarship in AAC has cautioned against equating access to a synthetic voice with meaningful self-expression. Alper \cite{alper2017giving} argues that AAC technologies are often framed as ''giving voice'' to users, yet the ability to communicate, be recognized, and be heard on one's own terms remains shaped by broader social, cultural, and economic factors. This perspective positions voice not merely as a technical artifact to be personalized, but as a sociotechnical resource whose value depends on how well it supports identity, agency, and participation in everyday life. \citet{preece2024making} extend this concern beyond technology, arguing that voice personalization involves ongoing questions of identity and agency: who controls one's voice, and what happens when that control is compromised. \citet{weinberg2025robot} expanded this to lexical voice through an autoethnographic deployment of a personalized LLM, finding that while deep personalization enhanced fluency and relevance, it also introduced new tensions around authorship and self-expression. 
 
Yet voice personalization remains difficult to specify. \citet{pullin201517of} demonstrate that people with complex communication needs have very little expressive control over tone of voice, and that there are ``17 ways to say yes'', subtle expressive distinctions that matter deeply for in-the-moment social interaction, but that current systems cannot capture. \citet{pullin2017designing} further argue that voice should be treated as something users co-create over time. Noufi et al. \cite{noufi2023role, noufi2023context} propose a model of vocal persona, the contextually-dependent, continuous expressive space that a voice inhabits, arguing that incorporating vocal persona into speech synthesis can significantly enhance the agency and sense of authenticity experienced by AAC users. Modern generative TTS models bring this goal technically closer, but specifying a culturally resonant voice through free-form prompts remains a barrier. Prior work shows that non-experts often struggle to translate intentions into prompts that AI systems can reliably interpret~\cite{ippolito2022creative, zamfirescu2023johnny}. This suggests that advances in generation alone are insufficient: meaningful voice personalization also requires accessible ways for users to articulate, explore, and refine identity-linked vocal preferences.
The bespoke, iterative nature of voice personalization we observe also connects to a broader tradition of DIY assistive technology, in which disabled people adapt and customize off-the-shelf tools to meet needs that commercial products fail to address~\cite{hurst2011empowering}.

\subsection{Identity Representation in AAC Systems}

Cultural and linguistic diversity is under-addressed in AAC design. \citet{baxter2012barriers} document that the voice and lexical expression of the device is a consistently cited barrier to AAC adoption. ~\citet{alper2017giving} argues that AAC voices are not merely communication outputs, but are intertwined with identity, recognition, and participation. The value of a voice depends not only on intelligibility, but also on whether it enables users to represent themselves in ways that reflect their authentic identities. ~\citet{michel2025bias} examined accent bias in AI voice cloning services (\textit{Speechify} and \textit{ElevenLabs}), finding that American and British accents dominate commercial offerings, and that users from non-dominant linguistic communities experience both lower output quality and a limited range of voices that feel representative, with participants describing available synthetic voices as \textit{``not a representation of me.''} 
\citet{terblanche2025you} found that children with expressive communication difficulties in South Africa preferred newly developed synthetic voices in their home languages (South African English, Afrikaans, and isiXhosa) over British or US English voices, demonstrating that stakeholders in under-resourced language communities actively seek voices that reflect their own linguistic and cultural backgrounds. Related work further shows that voice representational gaps extend beyond accent and language. \citet{danielescu2023creating} demonstrate how TTS encodes binary gender assumptions, demonstrating the need for synthetic voices that cover the full range of gender presentation.
 
These dynamics connect to deeper questions of self-recognition. \citet{payne_perceptual_2021} show experimentally that when people are given agency in \textit{choosing} a self-associated voice, that voice becomes more perceptually distinctive relative to others. Because stimuli associated with oneself are perceived as especially salient, this suggests that choosing a voice can strengthen identification with it, making voice selection more than a matter of preference.
For AAC users from non-dominant linguistic backgrounds, these dynamics compound by broader risks documented outside of AAC literature.  Related work further shows that generative AI systems can reshape lexical voice by steering user language toward standardized and normative forms (e.g., Standard American English), at times neutralizing the wording, tone, and stylistic distinctiveness of racialized dialects and ethnolects (e.g., African American Vernacular English) often used for self-expression~\cite{glazko2025autoethnographic,johnson2026don}. AI suggestions have been shown to homogenize expression toward dominant, Western styles and diminish cultural nuance~\cite{agarwal2025ai}. Similarly, Lawrence \cite{lawrence2021siri} argues that speech technologies often privilege dominant linguistic norms, creating barriers for speakers whose accents, dialects, or communication styles fall outside those expectations. Likewise, literature on speech technologies suggests that many systems are optimized around narrow acoustic ideals, often privileging dominant prosodic patterns \cite{blodgett2020language, koenecke2020racial}.  Further, LLMs have been shown to systematically misrecognize ableism in ways shaped by cultural context~\cite{phutane2025disability}. 
These dynamics are especially consequential in AAC, where the \textit{prosodic voice} of a device is the voice others associate with the user. An open question is how cultural voice (mis)alignment is experienced by AAC users from non-dominant linguistic backgrounds, and whether design approaches that address identity, belonging, and self-recognition can better support representation.

\section{Survey of AAC Users}

To characterize how well current synthetic voices represent users' cultural and linguistic identities, we conducted an online survey targeting AAC users across multiple gender identities, countries, and language communities. The survey was distributed in English and Spanish through AAC-focused online communities, including AAC users on different AAC users Facebook groups, the AccessComputing mailing list, advocacy organizations, and direct outreach to participants from prior studies conducted by our lab. We compensated participants 5 dollars each for completing the survey.

The survey collected demographic information (e.g., country of birth and residence, languages spoken, AAC device used) and assessed four dimensions: overall satisfaction with current voice options, perceived cultural and linguistic representation (1–5 Likert scale), specific sources of frustration (e.g., limited dialect options, mispronunciations, robotic quality, difficulty personalizing), and prior attempts at voice customization. 
Participants who had personalized their voice were asked to describe the process and rate its difficulty, while those who had not were asked to explain why. The survey closed with an open-ended question inviting additional reflections on their AAC voice or device. A full set of questions can be found in Appendix \ref{appendix:a}. The study protocol and data processing were approved by our institution's IRB STUDY00005427.

\subsection{Participant Verification} 

\begin{figure}
    \centering
    \includegraphics[width=1\linewidth]{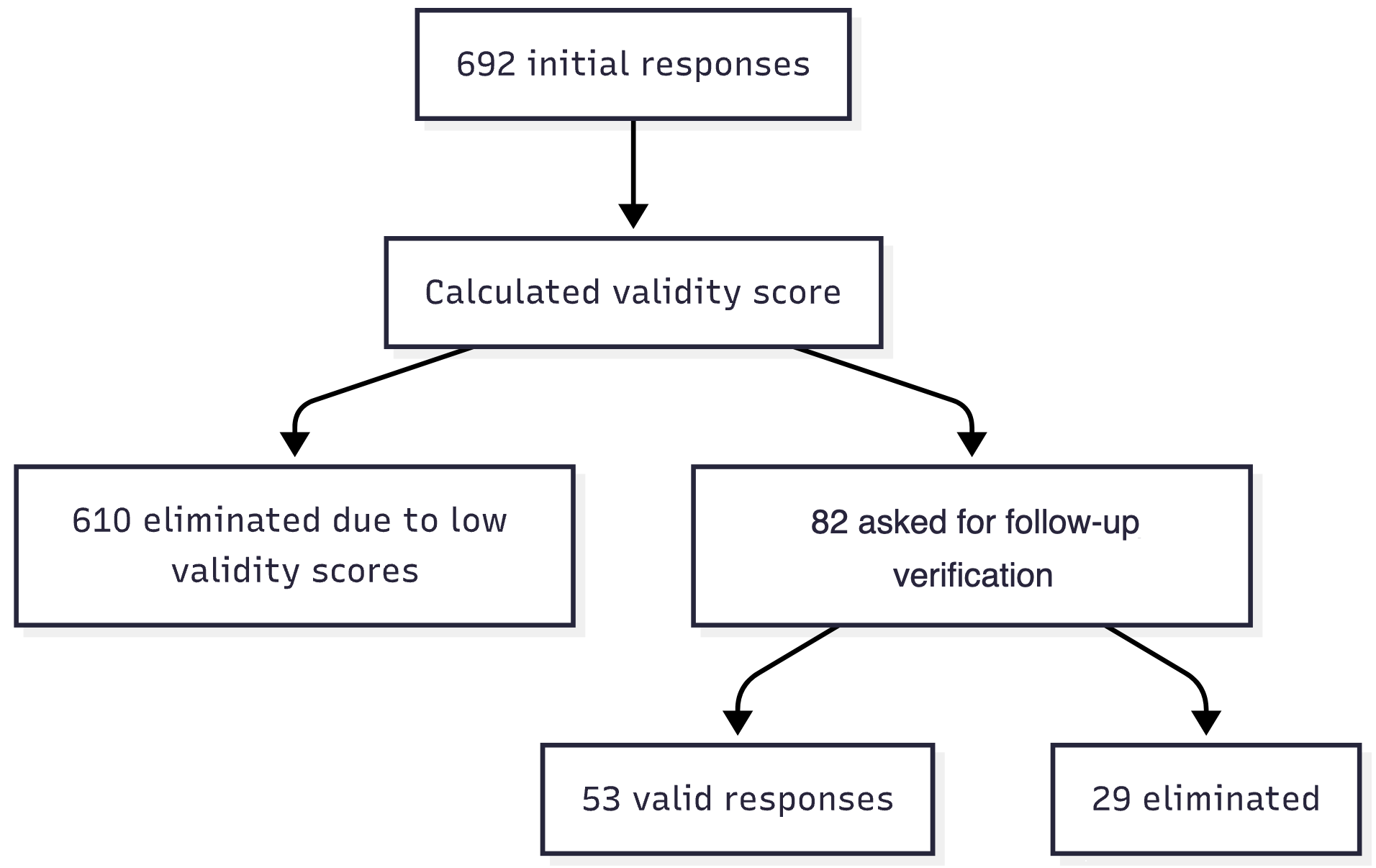}
        \caption{Diagram of survey response verification and exclusion pipeline.}
    \label{fig:validity}
    \Description{
flowchart TD
    A[692 initial responses]
    B[Calculated validity score]
    C[610 eliminated due to low validity scores]
    D[82 asked for follow-up verification]
    E[29 eliminated]
    F[53 valid responses]

    A --> B
    B --> D
    B --> C
    D --> F
    D --> E
    }
\end{figure}
Of 692 total responses received, 53 were retained for analysis. Recruiting AAC users through publicly accessible online surveys presents challenges around response authenticity. AAC users are a relatively small and hard-to-reach population, so we distributed the survey broadly. However, survey links with monetary compensation can attract responses from malicious actors. Many responses appeared to be from individuals outside the AAC community. Others were AI-generated, fluent, and topically relevant but lacking the specificity, personal detail, and linguistic idiosyncrasy characteristic of lived AAC experience. 
 Fig. ~\ref{fig:validity} shows our selection process, which is described next. 
 
Issues of survey integrity have been documented in survey research more broadly~\cite{bonnamy2025survey, irish2023bots}. To address this, 
we embedded several verification items directly in the survey instrument. We asked participants how they had heard about the survey (e.g., through a specific AAC organization, social media, or word of mouth), which served as a plausibility check against our actual recruitment channels. Critically, this item functioned as a trap question: responses naming recruitment channels we had not used served as a signal of inauthentic responding, regardless of whether the stated channel was plausible in general. We also included a multiple-choice question asking participants to identify which AAC-related organizations they were affiliated with or familiar with; this list included fictitious organization names alongside real ones. Respondents who selected fake organizations were flagged for exclusion, as it indicated either inattentive responding or unfamiliarity with the AAC community. 

Following initial collection, we constructed a metric to assess the likelihood that a response was valid. To construct this, we checked for internally inconsistent answers across screener items, recruitment plausibility, and selection of fictitious organizations. We also reviewed all open-text responses for coherence, specificity, and plausibility, flagging responses that appeared automated or implausible. This eliminated 610 responses. We sent a follow-up verification request to all 82 respondents not already excluded, asking them to provide one form of identity confirmation, such as a photograph of their AAC device, a link to an institutional page or personal website, proof of membership in an AAC organization, or any other evidence they felt appropriate. 
An additional 29 either did not respond or sent patently false documentation, such as images of famous AAC users, leaving 53 valid responses for further analysis.

Some legitimate AAC users may have been excluded if they were unable or unwilling to provide verification. This is a real concern given that the burden of proof falls disproportionately on a population that already faces significant barriers to participation in research. For the purpose of this work, we erred on the safe side to increase the validity of our findings. We report these limitations and note that the challenge of bot and fraudulent responses in accessibility survey research warrants broader methodological attention from the community. 

%
\begin{table}[h]

\begin{tabular}{llr}
\toprule
\textbf{Dimension} & \textbf{Category} & \textbf{\%} \\
\midrule
Gender & Man & 52.8\% \\
       & Woman & 32.1\% \\
       & Non-binary & 15.1\% \\
\midrule
Age & 18--24 & 9.4\% \\
    & 25--34 & 60.4\% \\
    & 35--44 & 26.4\% \\
    & 45--54 & 3.8\% \\
\midrule
Country of Birth & USA & 81.1\% \\
                 & Argentina & 7.5\% \\
                 & Other & 11.4\% \\
\bottomrule
\end{tabular}
\caption{Demographic distribution of the 53 validated survey respondents by gender, age, and country of birth.}
\label{tab:demographicstable}
\end{table}

\subsection{Analysis}
We used descriptive statistics and Mann-Whitney U tests to analyze Likert-scale responses, treating our ordinal data non-parametrically. For each two-group comparison, we report the rank-biserial correlation ($r_{\textit{rb}}$) as an effect size, where values of .1, .3, and .5 correspond roughly to large, medium, and small effects. We conducted two planned comparisons across all eleven Likert dimensions: non-binary respondents (n=8) versus men and women pooled (n=45), and respondents born outside the United States (n=10) versus those born in the United States (n=43). We report all tests we ran, including non-significant outcomes. For the intersection of these two axes, cell sizes were too small for inferential testing (n=1 for the combined non-binary and non-US-born group); we report those comparisons descriptively as an illustration of compounded disadvantage rather than as inferential claims. 
We thematically coded the open-ended responses and inductively derived the following themes: (1)~Current voices meet communicative needs but not expressive ones, (2)~Non-binary respondents are systematically under-served, (3)~Country of birth is a secondary axis.

\subsection{Findings}

\begin{figure}
    \centering
    \includegraphics[width=\linewidth, trim={0 25 0 25}, clip]{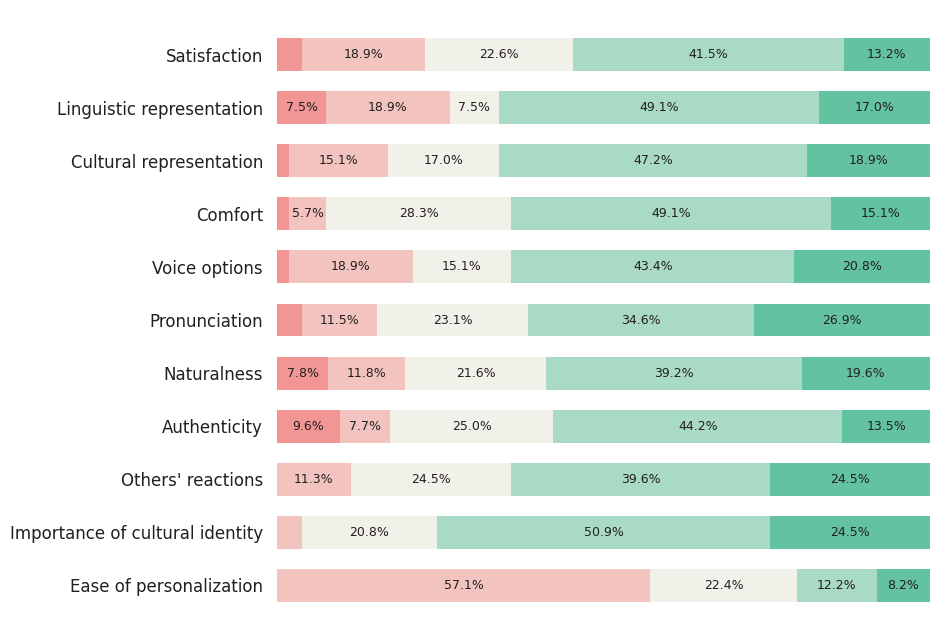}
    \caption{Distribution of responses across the eleven survey Likert measures.}
    \Description{A horizontal diverging bar chart showing Likert-scale response distributions (1–5, negative to positive) for 11 survey dimensions. Most dimensions show majority positive responses (4 or 5). Ease of personalization is the notable exception, with 57.1\% of responses rating it 1 or 2. Importance of cultural identity has the highest proportion of positive responses (75.4\% rating 4 or 5). Satisfaction, linguistic representation, cultural representation, comfort, voice options, pronunciation, naturalness, authenticity, and others' reactions all cluster slightly above the midpoint.}
    \label{fig:survey_likertscale}
\end{figure}

Of the 53 valid respondents,  about half identified as men, 1/3 women, and 15\% nonbinary, with ages ranging from 18 to 54, as summarized in Table \ref{tab:demographicstable}.  Most (81\%) participants were from the United States. Across the 53 respondents, current synthetic voices were rated as moderately supportive of communication but rarely as expressive of identity. As shown in Fig. \ref{fig:survey_likertscale} overall ratings clustered around slightly above the midpoint on every representational dimension: satisfaction $M{=}3.42$ ($SD{=}1.06$), linguistic representation $M{=}3.49$ ($SD{=}1.2$0), cultural representation $M{=}3.66$ ($SD{=}1.02$), comfort $M{=}3.70$ ($SD{=}0.87$), 75\% of respondents rated the importance of having their voice reflect their cultural identity a 4 or 5 on a five-point scale $M{=}3.96$ ($SD{=}0.78$). The gap between what participants wanted from their voice and what they had was not evenly distributed. Non-binary respondents (n=8) rated nearly every dimension approximately one point lower than men and women pooled (n=45), with statistically significant differences on most outcomes (e.g., satisfaction: M=2.62 vs. 3.56, p<.01; others' reactions: M=2.88 vs. 3.93, p<.05). Non-US respondents  (n=10) showed a similar pattern, rating overall satisfaction and cultural authenticity significantly lower than US-born respondents (satisfaction: M=2.70 vs. 3.58, p<.05; cultural authenticity: M=2.90 vs. 3.57, p<.05).

\subsubsection{Current Voices Meet Communicative Needs But Not Expressive Ones.}

Most participants described their device voice as functional for day-to-day communication but generic in tone and identity. 64\% rated the available voice options positive (4 or 5), and 64\% rated others' reactions to their voice positive, yet open-text responses returned repeatedly to specific expressive failures rather than to satisfaction. Respondents described mispronunciations of names and loan words (S2: \textit{``It is embarrassing and breaks down communication when my voice butchers words like `spiciest' egregiously and I have to try and correct miscommunications while in conversation''}), the absence of prosody (S5: \textit{``Inability to change tone of phrase (such as to use sarcasm)''}), and mismatches between the voice and the speaker's age or personality (S33: \textit{``...it doesn't really sound like me. It feels too generic and doesn't match my age or personality''}). Multilingual respondents described being effectively monolingual on their device. S14 wrote, \textit{``It is not multilingual. I am not able to type in or use both languages. I'm stuck in English or heavily modifying the pronunciation for any Irish language buttons I add.''} Spanish-speaking respondents raised the same complaint: S52 noted \textit{``Las voces en español, son de España y no latinas'' (The Spanish voices are from Spain, not Latin America)}, and S51 said \textit{``no hay Argentino'' (there is no Argentinian)}. 

\subsubsection{Non-binary Respondents are Systematically Under-served.}

\begin{figure}
    \centering
    \includegraphics[width=1\linewidth]{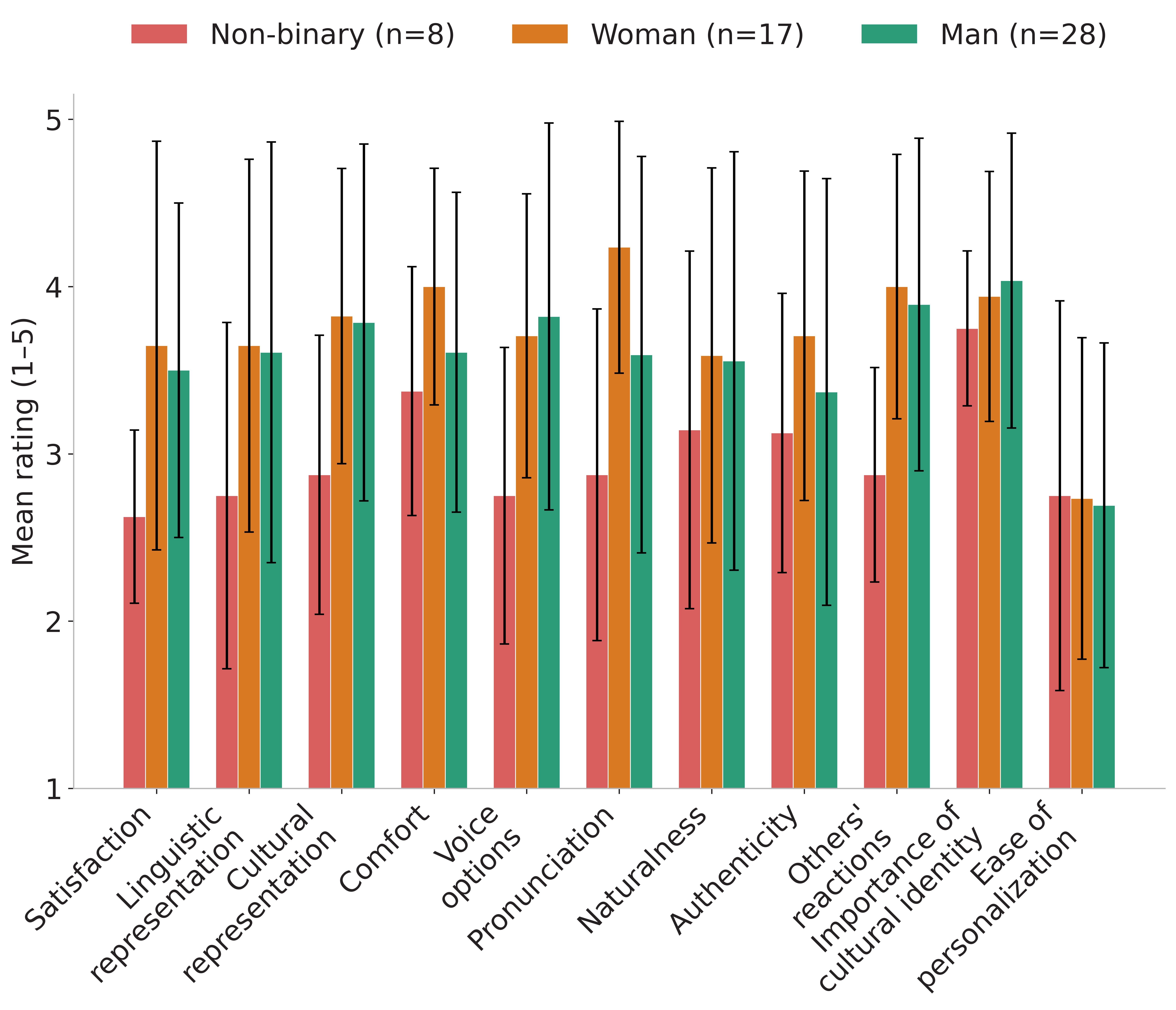}
    \caption{Mean ratings (1–5) across survey dimensions comparing non-binary~(n=8), men~(n=28) and women~(n=17) AAC users pooled.}
    \Description{Grouped bar chart showing mean Likert ratings from 1 to 5 across eleven dimensions: Satisfaction, Linguistic representation, Cultural representation, Comfort, Voice options, Pronunciation, Naturalness, Authenticity, Others' reactions, Importance of cultural reflection, and Ease of personalization. Each dimension shows two bars — coral pink for non-binary users (n=8) and teal for men and women pooled (n=45) — with error bars representing one standard deviation. Non-binary users score lower than men and women on all dimensions except Ease of personalization, where means are nearly identical (2.75 vs. 2.71). The largest gaps appear on Others' reactions (2.88 vs. 3.93, p<.01), Satisfaction (2.62 vs. 3.56, p<.01), Voice options (2.75 vs. 3.78, p<.05), and Cultural representation (2.88 vs. 3.80, p<.05). Linguistic representation (2.75 vs. 3.62) and Comfort (3.38 vs. 3.76) are marginally significant (p<.10). Naturalness and Authenticity show no significant difference.}
    \label{fig:survey_gender}
\end{figure}

The data also revealed a strong effect related to gender identity alignment. Non-binary respondents ($n{=}8$) rated nearly every dimension about one point lower than men and women pooled ($n{=}45$), with medium-to-large effect sizes and statistically significant differences on most outcomes, as we can see in Fig. \ref{fig:survey_gender}.
The open-ended responses further reinforce these quantitative findings. Commercial voice libraries present voices in two buckets, male and female, and non-binary respondents described being forced to: (1) choose between the two, (2) accept a child voice as the closest proxy to a gender neutral voice, or (3) use the voice as a deliberate form of mismatch. S5 wrote, \textit{``The options are either very feminine, very masculine, or children. There aren't any gender neutral voices or any more feminine men's voices or `gay' sounding men.''} S8 expressed that \textit{``People will give me a hard time about using a child's voice, but it's the best option I have.''} And S13 said \textit{``I don't actually have `sounds like a human' as a priority for my AAC voices, and I'd kind of like my voices to confuse people about my gender. \textit{Voice options} and \textit{Pronunciation} are problems whenever I'm actually acting like the multilingual I am.''} S6, an Australian non-binary respondent, described the same gap alongside a desire for regional specificity: \textit{``I'd love a gender neutral voice that fits an early 20-year-old from my city... I'm from a very bogan (lower class) family, and I would love it if I could swear and say certain phrases with the right twang to them.''} 

We read this as evidence that the dominant engineering framing of voice selection as a binary gender choice is itself the constraint.

\subsubsection{Country of Birth is a Secondary Axis.}
Splitting by country of birth (non-US vs US) also revealed a strongly significant effect. Respondents born outside the United States ($n{=}10$) rated overall satisfaction significantly lower than US-born respondents ($n{=}43$) ($M{=}2.70$ vs.~$3.58$; $U{=}131$, $p{<}.05$, $r_{\textit{rb}}{=}.39$), and also rated their voice's cultural authenticity significantly lower ($M{=}2.90$ vs.~$3.57$; $U{=}129$, $p{<}.05$, $r_{\textit{rb}}{=}.39$). The open-ended responses from non-US-born participants mirror \citet{terblanche2025you, terblanche2025development}'s observation that stakeholders in under-resourced linguistic communities actively seek voices in their own varieties: 

S52 wrote \textit{``No hay voces mexicanas'' (there are no Mexican voices)}; S53 wrote \textit{``Me representa porque la vengo usando hace más de la mitad de mi vida. Pero no representa a mi acento'' (It represents me because I've been using it for more than half my life, but it does not represent my accent)}. 

\subsubsection{Personalization is the Shared Frustration.}
Personalization is where every group agrees. Ease of personalization was rated $M{=}2.71$ (median $=2$), with 57\% of respondents rating it negative (1 or 2  on a five-point scale), and we found no significant difference by gender or country of birth; this was a shared experience of friction. The open-ended responses contextualize this. S6, who had tried Apple's Personal Voice, described the recording workload as inaccessible: \textit{``I have tried a personal voice from Apple, it was difficult to complete all the recordings and the result sounded very American, I've never used it because it makes me viscerally uncomfortable and there aren't any financially accessible custom voice options for me so I've stuck with a non-custom one.''} S12, who had manually corrected pronunciation, described the work as continuous and incomplete: \textit{``I have to go in and change each word that it pronounces differently from how I did, and this is time-consuming and always incomplete.''} Notably, respondents who did describe a positive personalization experience framed the outcome in affective rather than technical terms. For example, S20 wrote \textit{``I customized the pitch and tone with my speech therapist, so it matches my old voice recordings before my stroke. It was emotional but worth it,''} consistent with \citet{payne_perceptual_2021} on how the act of choosing a voice changes the user's relationship to it. 
\section{Qualitative Interviews: AAC Voices Grow on Users with Time, but Are Rarely Fully One's Own}

We conducted qualitative interviews with six AAC users to understand how AAC users communicate the characteristics of their ideal voice, and what it means for a voice to feel culturally aligned. Our secondary goal is to evaluate how users reflect on new culturally aligned voices. To this end, we generate a voice for participants based on their background and ask them to reflect on that. The interview protocol and data processing were approved by our institution's IRB STUDY00005427.

\subsection{Interview Study Participants}

\begin{table*}[h]
    \centering
    \small
    \renewcommand{\arraystretch}{1.1}
    \begin{tabular}{lllllll}
         \textbf{ID} & \textbf{Age} & \textbf{Gender} & \textbf{Race} & \textbf{Ethnicity} & \textbf{Cultural identity} & \textbf{Language(s) spoken} \\
        
         P1 & 25-34 & Non-binary & White/Caucasian & Not Hispanic/Latino(a) & Ashkenazi Jewish & \begin{tabular}{@{}l@{}}American English, \\Mandarin Chinese, and \\small but culturally \\meaningful amounts of \\Yiddish + Hebrew\end{tabular}\\
         P2 & 45-54 & Male & White/Caucasian & Not Hispanic/Latino(a) &  & English \\
         P3 & 25-34 & Female & White/Caucasian & Hispanic/Latino(a) &  & English \\
         P4 & 35-44 & Male & Black or African American & Not Hispanic/Latino(a) &  & English \\
         P5 & 25-34 & Male & \begin{tabular}{@{}l@{}}Hispanic or Latino or \\Spanish Origin\end{tabular} & Hispanic/Latino(a) & Argentine & Spanish \\
         P6 & 25-34 & Male & White/Caucasian & Hispanic/Latino(a) & Argentine & \begin{tabular}{@{}l@{}}Spanish, Portuguese, \\(basic) Italian \end{tabular}\\
    \end{tabular}
    \caption{Participant demographic details.}
    \Description{This table presents details of six participants, including their age, gender, race, ethnicity, cultural identity, and languages spoken.}
    \label{tab:participants1}
\end{table*}

We recruited six AAC users from the pool of survey respondents who had indicated willingness to participate in follow-up research. Participants self-described the source of their speech disability as: cerebral palsy, head trauma, a neuromuscular condition, and being on the autism spectrum. We required participants to be adults (18+) who currently use a speech-generating device as a primary or significant mode of communication, and who self-identify as having a culturally underrepresented or multilingual background. Demographics are shown in Table~\ref{tab:participants1}.

Participants disclosed a range of severity in their motor disabilities; we grouped them into the following categories: Mild (minor limitations in movement, 2 participants), Moderate (noticeable limitations, performs daily activities with some assistance, 3 participants), and Profound (unable to perform daily activities without significant assistance, 1 participant). 
Participants used different types of input modalities for their AAC, including direct touch with and without a keyguard, a keyboard, eye gaze, and switch access. AAC devices included a laptop, Tobii Dynavox, Accent1400, NuVOICE, and Proloquo4Text.

\subsection{Interview Study Procedure}
We conducted semi-structured interviews over Zoom. Audio and video were recorded and transcribed using Zoom Cloud. Each study session lasted two to four hours to accommodate participants' speaking rate and consisted of three phases, which we detail below: (1)~manual prompt creation, (2)~interviewer-tool-based prompt generation, and (3)~a comparative reflection interview. Interviews were conducted by the lead author, who is a long-term AAC user, as disclosed in the positionality statement, adding an additional layer of perspective and trust during interviews. All six participants responded using their AAC device throughout the interview. 

\subsubsection{Task 1: Manually Define ``Your'' Voice. }
The first task explores how participants, on their own, come up with their preferred voice. It provides a perspective of how participants reason about their ideal voice on their own.

We introduced participants to \textit{ElevenLabs}, an AI voice synthesis platform which generates prosodic voices from free-text descriptions in the form of ``prompts''. To ground participants in what such a prompt could look like and how it translates into a synthetic voice, we demonstrated a brief example, ``A 30-year-old woman with a calm tone and a slight Caribbean accent'', we generated the resulting prosodic voice, and played it. We asked participants to describe their ``ideal'' AAC voice, and to write a prompt in their own words. Based on that prompt, we generated three voices in \textit{ElevenLabs} and asked participants to choose their preferred voice. To help them evaluate the voice in context, participants composed four short sentences they might say in everyday situations (to a close friend, a coworker, or a family member), and we played those sentences aloud using the selected voice.
Participants completed an evaluation form (Appendix \ref{appendix:b}) rating the synthetic voice (using a 5-point Likert scale) on dimensions including cultural and linguistic representation, naturalness, accent accuracy, and social comfort.

\subsubsection{Task 2: Generate a Synthetic Voice Based on a Cultural Survey.} 

Based on the lead author's personal experience shaping his Argentine synthetic AI voice, we hypothesized that participants would have a hard time defining their own culturally aligned voice. To explore how to support this process, we built a simple tool that generates voice prompts based on a cultural survey consisting of 10 dynamically generated probing questions covering topics like culture, language, region, and speech style. We note that this study was not designed as a controlled evaluation of the probing tool; we do not claim that Voice 2 is statistically superior to Voice 1. The voice generation process serves as an elicitation mechanism, a way to surface how AAC users articulate, reflect on, and negotiate cultural voice identity, rather than to benchmark voice quality.
 
Participants reviewed the generated prompt and were invited to suggest modifications before synthesis. As in Task~1, three candidate voices were generated; participants selected one, listened to it reading the same sentences from Task~1, and completed the same evaluation form.

\paragraph{\textbf{Probing Tool}}

\begin{figure}[h]
    \centering
    \includegraphics[width=0.95\linewidth]{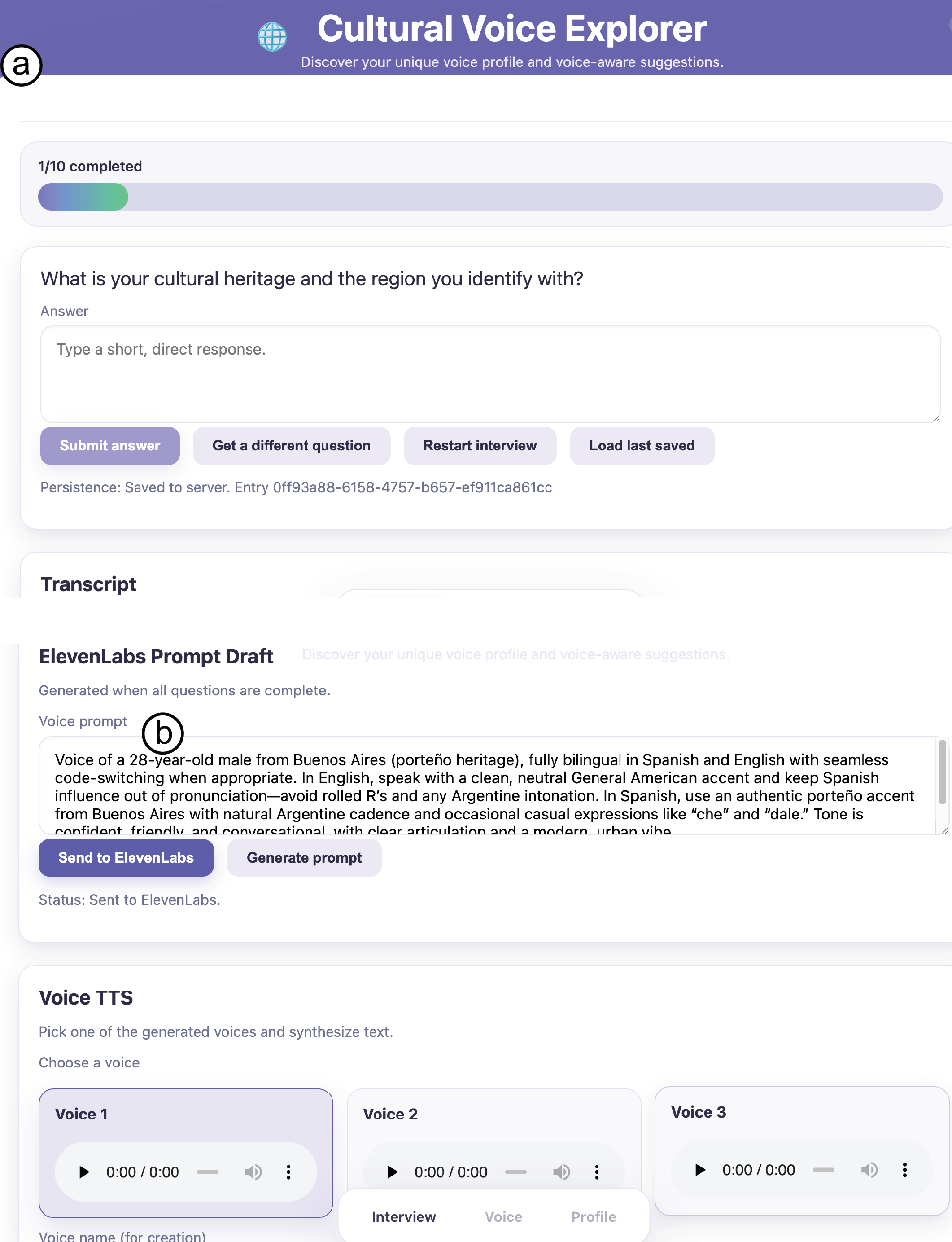}
    \caption{The cultural probing tool used in Task 2. (a) Participants progress through 10 dynamically generated questions about their cultural identity.  (b) The tool converts those responses into a personalized ElevenLabs prompt and generates three candidate voices for selection.}
    \Description{Two screenshots of the Cultural Voice Explorer web application on a purple background. (a) shows the active interview interface: a progress bar indicating 1 of 10 questions completed, the question "What is your cultural heritage and the region you identify with?" with a text input area below, four action buttons (Submit answer, Get a different question, Restart interview, Load last saved), a persistence confirmation message. (b) shows the Voice tab with a generated ElevenLabs prompt describing a bilingual speaker with a porteño Spanish accent and neutral General American English accent. Below the prompt are a "Send to ElevenLabs" button, a "Generate prompt" button, a status indicator reading "Sent to ElevenLabs," and three unlabeled audio players labeled Voice 1, Voice 2, and Voice 3 for candidate voice selection.}
    \label{fig:tool}
   
\end{figure}

We developed a simple probing tool to survey participants on cultural markers. The tool then composes an aligned prompt for \textit{ElevenLabs}. Fig. \ref{fig:tool}a illustrates how the probing tool  
prompts users with 10 questions about their cultural and linguistic background. Users can request a different question if they do not know how to answer a question or if they feel the question does not highlight a relevant dimension of their background. The questions are dynamically generated by a GPT-4o-mini model, prompted to explore users' multicultural identity (full prompt in Appendix \ref{appendix:c}). In this way, questions are not predefined, and they dynamically adapt based on the user responses.

Once the questions are answered, another GPT-4o-mini uses the transcript to create a voice prompt (Fig. \ref{fig:tool}b), and the user can send the prompt to \textit{ElevenLabs} to get the three voice candidates or modify the prompt if desired.

\subsubsection{Task 3: Reflection.} 
To understand how participants reflect and respond to the generated voices and the cultural and/or emotional alignment of these, we concluded the study with an in-depth semi-structured interview.

We asked which voice felt more like theirs and why, whether the probing tool surfaced aspects of their cultural identity they might not have otherwise articulated, moments when a voice captured or failed to capture something meaningful, their sense of agency and authorship over the generated voices, concerns about stereotyping or misrepresentation, and contexts in which they would or would not feel comfortable using the voice. Here, we restate that the lived experience of the interviewer plays a role in terms of creating trust, but also possibly subconsciously steering the conversation. 

\subsection{Analysis}
We analyzed interview transcripts using inductive coding followed by reflective thematic analysis~\cite{clarke2017thematic}, with codes grouped through affinity diagramming into the four themes reported below.
Voice evaluation forms collected after each task used a 5-point Likert scale across ten dimensions: cultural match, linguistic identity, pronunciation accuracy, accent/regional fit, identity alignment (whether the voice felt like a version of their own), naturalness, tonal fit, social comfort, sense of representation, and stereotyping (reverse-coded). Given the small sample size (n=6), we analyze these descriptively and treat ratings as contextual rather than generalizable. Finally, we collected and compared the prompts participants authored in Task 1 against the tool-generated prompts from Task 2, treating these as artifacts that reveal how participants conceptualize and externalize their vocal identity.

\subsection{Results}
\begin{figure}[h]
    \centering
    \includegraphics[width=\linewidth]{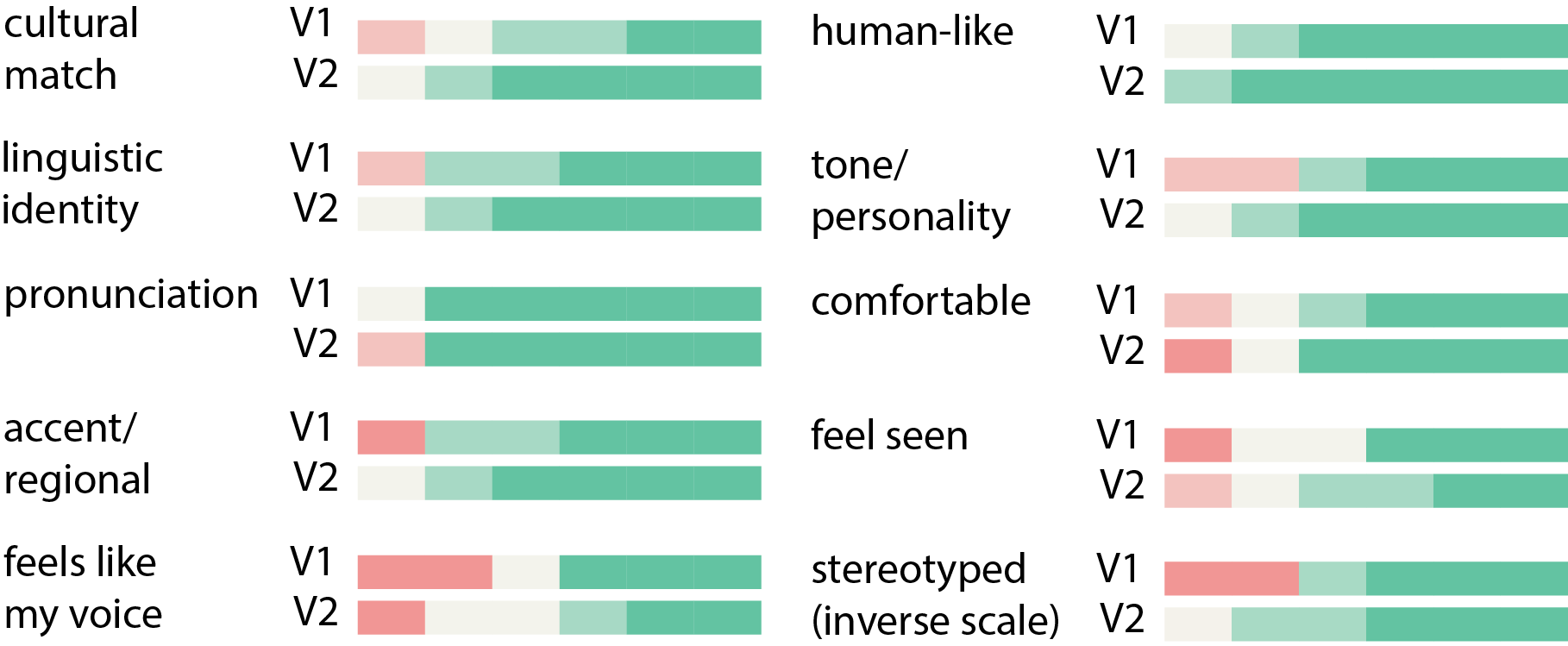}
    \caption{Participant-level evaluations of Voice 1 and Voice 2 across ten voice dimensions.}
    \Description{A grid of horizontal diverging bar charts comparing evaluation ratings for Voice 1 (manually authored prompt, V1) and Voice 2 (tool-generated prompt, V2) across ten dimensions: cultural match, linguistic identity, pronunciation, accent/regional fit, feels like my voice, human-like, tone/personality, comfortable, feel seen, and stereotyped (inverse scale). V2 shows higher (more positive) ratings than V1 on most dimensions, with the most notable improvement in cultural match. The stereotyped dimension shows greater consistency in V2. The feels-like-my-voice dimension shows the smallest improvement between V1 and V2."
Table 1: "Participant demographic table for the six qualitative interview study participants. Columns are: ID, Age, Gender, Race, Ethnicity, Cultural identity, and Languages spoken. P1: 25–34, Non-binary, White/Caucasian, Not Hispanic/Latino(a), Ashkenazi Jewish American, English, Mandarin Chinese, Yiddish, Hebrew. P2: 45–54, Male, White/Caucasian, Not Hispanic/Latino(a), English, English. P3: 25–34, Female, White/Caucasian, Hispanic/Latino(a), English, English. P4: 35–44, Male, Black or African American, Not Hispanic/Latino(a), English, English. P5: 25–34, Male, Hispanic or Latino or Spanish Origin, Hispanic/Latino(a), Argentine Spanish, Argentine Spanish. P6: 25–34, Male, White/Caucasian, Hispanic/Latino(a), Argentine Spanish, Argentine Spanish, Portuguese, and basic Italian.}
    \label{fig:eval_data}
\end{figure}

 Across most dimensions, the voice generated by our tool (V2) received higher mean scores than the manually constructed voice (V1),  with the most notable improvements in cultural match as shown in Fig. \ref{fig:eval_data}, suggesting that the probing tool produced prompts that were more culturally aligned and less prone to unwanted stereotyping. The stereotyped dimension also became considerably more consistent in V2, indicating that the tool prompt reduced not just stereotyping on average but its variance across participants.
The dimension with the highest variance in both voices was ``this voice feels like a version of my own voice'' (V1: M=3.33, SD=1.97; V2: M=3.50, SD=1.52), which also showed the smallest improvement between voices. 

\subsection{Qualitative findings}

Our thematic analysis revealed the following themes: (1)~``Voice as Iterative Construction''; examining what happened when participants actively engaged in building and refining their voice through prompting and reflection, (2)~``Voice as Private Self'', capturing the persistent gap between participants' inner voices and the synthetic voices their devices project to the world; (3)~``Voice as Temporal Identity'', reflecting how AAC voices are assigned at a fixed moment in life and fail to grow alongside their users; and (4)~``Voice as Cultural Negotiation'', describing the complex and often fraught process of articulating cultural identity through voice synthesis;  examining what happened when participants actively engaged in building and refining their voice through prompting and reflection.

\subsubsection{Voice as Iterative Construction}

Participants had rarely, if ever, been asked to describe their ideal voice, and the prompting process itself became a site of discovery. How participants constructed, revised, and ultimately recognized their voice revealed as much as the voices themselves.

Task 1 exposed a tension between what participants knew about their voice, what they could describe, and what they wanted to sound like. P4 produced the shortest prompt, \textit{``African-American man with a southern style voice''}, not because his identity was simple, but because prompt crafting may require AI literacy and initiative that users do not necessarily have. P3 opened with \textit{``I would like to create a Hungarian or an Irish voice,''} then shifted to request a Scottish accent, and eventually wondered whether she wanted to sound like a New Yorker. P1, by contrast, produced the most linguistically precise prompt of any participant, specifying not just accent and language but the exact phonetic behavior they expected at the boundaries between American English, Mandarin Chinese, Hebrew, and Yiddish. 

Iteration revealed what the first attempts concealed. P5's first manual prompt described \textit{``una voz argentina de un joven canchero y un toque sarcástico, la voz suena semi grave, suave y seductora'' (``an Argentine voice of a cool young guy with a sarcastic touch, the voice sounds semi-deep, soft and seductive'')}. His second prompt removed \textit{``semi grave''} and \textit{``seductora''}, trimming descriptors he felt were misdirecting the voice generation software. P6 iterated across three manual prompts, each narrowing the regional specification: from \textit{``Rioplatense''} to \textit{``del interior de la provincia de Buenos Aires''} to \textit{``del interior de la provincia de Buenos Aires, Argentina.''} Each revision added geographic precision in response to voices that had defaulted to a Buenos Aires city accent. Through this process, he learned what the voice generation software understood and pushed back against its defaults.

The tool-generated prompts showed a different kind of iteration. P6's tool prompt went through two versions: the first described a \textit{``voz porteña de Buenos Aires'' (``porteño voice from Buenos Aires'')}, which the second revised to simply \textit{``de Buenos Aires,''} removing \textit{``porteña''} after P6 noted that the term was pulling the voice toward the urban accent he wanted to avoid. The revision was small but pointed: the adjective generated a regional stereotype, removing it was an attempt to create space for something more specific.

For P2, iteration looked different again. He produced one manual prompt and did not revise it, but his wife was present throughout the session, and her reactions to each candidate's voice, visible, immediate, and consequential, functioned as a form of collaborative iteration that the prompting interface had not been designed to accommodate. Their eventual choice was not his alone. It emerged from a negotiation conducted largely in glances, and the voice they selected was, in this sense, authored relationally.

Both P2 and P4 independently described the voice they had chosen as resembling a family member. P2 said it \textit{``sounded like his brother''}; P4 reflected that the tool-generated voice reminded him of his stepfather, who had passed away: \textit{``it's his voice... that's why I picked it''}. This recognition was ultimately why he chose it. Neither had mentioned family resemblance during manual construction, and neither had been asked about it in the survey. The recognition came only retrospectively, as the participants reflected on the different prosodic voices that were generated.

\subsubsection{Voice as Private Self}

Almost all participants (P1, P4, P5, P6) described a persistent gap between their \textit{``inner voice,''} the one in which their thoughts are narrated, and the synthetic voice their device projects to the world. This inner voice, vivid and specific to each participant, remained largely inaccessible to others, and as it turned out, extremely difficult to externalize.
P6: \textit{``my voice is my best-kept secret''}. He described living with two voices simultaneously: \textit{``Rodrigo,''} a Mexican Spanish TTS voice the world had come to know him through, and a different Argentine inner voice only he perceives. In Task 2, our tool generated Argentine voice candidates, closer to his identity than ``Rodrigo'', he spontaneously sent audio clips to close friends. Their responses were telling, \textit{``ese no sos vos'' (``that's not you'')}, revealing that his social circle had become so attached to ``Rodrigo'' that they could not recognize him in a different voice. TTS voices grow on users over time. ``Rodrigo'' had been P6's public voice for so long that his friends could not recognize him in a voice that reflected his culture.
P5 had internalized a similar displacement. Also Argentine, he had spent years communicating through \textit{``Antonio,''} a Castilian Spanish voice with a marked Gallego accent. He struggled to self-identify with the Argentine voices, not because they were wrong, but because ``Antonio'' had become so familiar that anything else felt off. \textit{``Es tipo… escucho un segundo y se va'' (``it's like… I hear it for a second and it's gone'')}, he said of his inner voice, suggesting that years of living through ``Antonio'' had not erased it, but had made the distance between inner and outer voice feel natural.

For P4, the gap between inner and outer voice carried additional weight. As a Black man whose device produced what he described as sounding like \textit{``a white man and a robot,''} the mismatch was a form of racial misrepresentation, his voice communicating an identity that was not his. With his current synthetic voice, P4 does not sound like a Black man, or anything that maps onto who he is. 

P1, too, encountered misrepresentation. Despite explicitly requesting a gender-neutral voice, the voice generation software consistently produced gendered output. They observed that the software \textit{``seems to be ignoring the gender-neutral thing''} and later, with more frustration, \textit{``my name is (female name), I am not a dude.''} This was not caused by poor prompt quality or insufficient detail, but by a synthesis engine that had no model of non-binary voice to draw from. This is the prompt they produced:\footnote{P1's original prompts contained Mandarin Chinese characters and Hebrew script. 
These are reproduced here in romanization.} 

\begin{quote}\small
A gender-neutral speaker of both American English and Mandarin Chinese. Their first language is English, spoken with a Boston accent. When the `r' sound is made rather than elided, it is a front of the mouth `r' similar to the Mandarin `er.' They also pronounce loanwords from Chinese, Hebrew, and Yiddish according to their source pronunciations and, where relevant, inflectional morphology. For example, ``Shanghai'' is always pronounced like [Sh\`angh\v{a}i], they pronounce the hard ``H'' in Hanukkah, and they will refer to multiple menorot rather than multiple menorahs. There is no other sign of a Chinese accent in their English. They speak standard Mandarin Chinese with a slight Beijing accent, including erhua. They regularly use multiple languages within a single sentence, including to say ``Z\v{e}nme f**king le'' [Mandarin/English mix] or to mention the menorot [Heb.] they saw at the Shanghai Jewish museum.
\end{quote}

The speech generation software had no notion of gender neutrality. P1 also pointed to a second layer of identity the system had no framework to represent: \textit{``none of them sounded autistic,''}  neuro-divergent speech patterns carry their own signature that they recognize as part of how they hear themselves, and that no generated voice came close to capturing. 

\subsubsection{Voice as Temporal Identity}
Unlike a natural voice, which evolves alongside a person, an AAC voice is assigned at a specific moment, often in childhood or adolescence, and remains fixed. The inner voice does continue to evolve as the AAC user ages. This creates a temporal mismatch that compounds over years.
For several participants (P2, P4, P5, P6), the voice they used was not one they had chosen so much as one they had been given, assigned at a particular moment in their life and carried forward unchanged while they continued to grow. 

P4 stated that at 36 years old, he had been using AAC since childhood, and the voices available to him throughout that time had never reflected who he was as a Black African American man. \textit{``For 31 years, I never heard a voice synthesizer sound like me,''} he said, a statement that lands with particular weight when considered against the broader history of assistive technology being designed by and for a predominantly white user base. 

P2, a white man from Pennsylvania with no strong claims to a particular cultural accent, had fewer axes of misrepresentation to navigate than P4. Yet when he heard a voice that felt fluid and human, his reference changed: \textit{``I do like the voice better than the couple of voices I had in high school.''} The comparison was not to an ideal or to his inner voice, but to the voices he got as a teenager. 

P5 and P6 both adopted AAC during their adolescence, meaning they had spent their adult communicative lives speaking through voices that were chosen as their identities were forming. P5 recalled using low-tech AAC as a child before acquiring ``Antonio'', his Castilian Spanish voice, in high school. \textit{``Desde el secundario, no use cuando era chico'' (``since high school, I didn't use it as a child'')}, he said, noting that he had carried an adult voice with a foreign accent since he was a teenager. P6 explored a series of TTS voices, ``Jorge'', ``Carlos'', and ``Diego'' from Loquendo, a Spanish woman, before settling on ``Rodrigo'', a Mexican neutral voice, in his late teens. \textit{``A los 8 cuando apareció Loquendo empecé a buscar esa voz'' (``at around 8 when Loquendo appeared I started looking for that voice'')}, he described a decade-long search that resolved because ``Rodrigo'' had accumulated enough life history to feel like his own. \textit{``Esa voz de Rodrigo ya era mi voz. Ya se había vuelto parte de mi identidad'' (``that voice of Rodrigo had already become my voice, it had already become part of my identity'')}.

What P5 and P6's histories reveal is a different kind of temporal mismatch than P4's. Where P4's voice never grew into him at all, P5 and P6's voices grew into them sideways, accumulating social weight and professional history without resolving the underlying cultural distance. P6 had built a public career, becoming what he described as the first mute person to do radio, through ``Rodrigo's'' voice. The voice had become inseparable from his professional identity even as it remained misaligned with who he was.

\subsubsection{Voice as Cultural Negotiation}
Articulating cultural voice identity, whether through a manual prompt or a probing tool, involves navigating layers of competing markers, regional specificity, linguistic history, and personal experience that current synthesis systems are ill-equipped to handle.

P1 presented the most complex case. As a non-binary, multilingual, neuro-divergent person who moves fluidly between American English, Mandarin Chinese, Hebrew, and Yiddish, their manual prompt was precise. However, the generated voices flattened that complexity. Chinese segments were skipped, \textit{``it skipped the Chinese!''}, and when they did appear, the boundary between languages collapsed in the wrong direction, English output picked up a Chinese accent P1 had not requested. \textit{``The first voices had Chinese accents in English. I am not Chinese,''} they said, pointing out the system's inability to hold multiple linguistic identities, which produced a racial stereotype. Culturally specific words fared no better: \textit{``AI fucked up the Hebrew,''} \textit{``also fucked up the Yiddish,''} and a mispronunciation of jiaozi as \textit{``yaoza''} which P1 flagged immediately. Mispronouncing identity-salient words, they suggested, is more damaging than generic errors because it signals that cultural modeling through the voice generation is superficial.
The negotiation revealed an implicit hierarchy. \textit{``Jewish beats Chinese for accenting the English,''} P1 observed, noting that when multiple cultural markers competed for influence over their English output, generated voices resolved the conflict by privileging one over the other rather than balancing diversity.

P5, Argentine, wanted a voice that could sustain a consistent regional accent throughout a sentence, something that felt coherent and settled. What he heard unpredictably fluctuated between Argentine and neutral within a single sentence, \textit{``y te vuelve al acento neutro'' (``and it goes back to the neutral accent'')}, a voice that could not commit to an identity. 
On the flip side, when P4 heard the generated voice, his reaction was immediate and visible, a smile that made this whole project feel worth it. Later, reflecting on what the tool's questions had surfaced, he said simply: \textit{``my heritage means a lot to me.''} For P4, the negotiation was less about competing markers or regional ambiguity and more about having, for the first time, a process that asked the right questions and having a system that generates a voice more aligned with his self-identity.

The stories highlighted so far included participants who had clear hopes and wishes for their voices. P3 struggled to identify what she wanted. When pushed to elaborate on what cultural identity meant to her in the context of voice, she said: \textit{``I'm stumped. That's going to be a margin of error, I feel like, because everybody is so mixed nowadays.''}
What made P3's case revealing was a moment later in the session when, after agreeing that the generated voices aligned well with her identity, she admitted: \textit{``I had to give you something.''} This suggested a limitation of the task itself: it assumed participants had a retrievable cultural voice identity to articulate. She had navigated the prompting process by selecting something plausible and going along with it. Her experience serves as a counterpoint, revealing that the negotiation of cultural voice identity presupposes a legible cultural identity to negotiate from, and that for some users, that foundation is genuinely uncertain.

\section{Discussion}
This work set out to understand how well current synthetic voices represent AAC users' cultural and linguistic identities, and then to understand what voice alignment means to those users in practice. Across our survey and interviews, what participants wanted from a voice went beyond accent accuracy or linguistic correctness: it encompassed self-recognition, belonging, and a felt sense of ownership. 
In the following sections, we reflect on what these findings imply for how identity alignment should be measured, how expanding the representational range of synthesis systems is central to serving users whose identities current models were not built to generate, and how understanding prosodic voice as a form of device-mediated identity reframes what meaningful voice personalization should build in the future.

\subsection{Beyond Cultural Accuracy: Measuring What Alignment Means}
One tension running through both studies is the difficulty of measuring what we care about. Our survey captures satisfaction, representation, and ease of personalization across eleven Likert scales, and our qualitative evaluation form asks participants to rate cultural match, linguistic identity, and tonal fit. These are reasonable proxies, but the dissociation we observed between cultural match scores and identity alignment scores in the qualitative interview study points to something they cannot fully capture. A voice can score well on cultural accuracy, accent, and regional specificity, while still failing to feel like one's own. High-fidelity voice cloning complicates this further: users who have banked their own voice have reported that the resulting synthetic voice crosses into an uncanny valley, recognizably theirs in acoustic properties but wrong in ways that are difficult to articulate, suggesting that identity alignment is not simply a function of acoustic similarity to one's own voice. 
The gap reflects a deeper measurement problem. Identity alignment in voice is not a property of the voice alone but of the relationship between a voice, a person, and the social history that voice has accumulated.~\citet{payne_perceptual_2021} showed that the act of selecting a self-associated voice changes the user's perceptual relationship to it; ownership itself shapes alignment, independently of the voice's acoustic properties. This means that any snapshot measure of identity alignment, taken at the moment of first exposure, will underestimate alignment that develops through use.
A single session cannot distinguish between a voice that feels right and a voice that has not yet had time to feel wrong.

This has methodological implications for the field. Studies that evaluate voice personalization through immediate ratings after brief exposure, including our own, are measuring something real but partial. Building better measures of identity alignment, ones that are sensitive to temporal development, social context, and the difference between cultural accuracy and felt ownership, is a methodological precondition for the field to evaluate whether voice personalization tools are actually doing what they claim.
Longitudinal diary methods, repeated short evaluations spaced across weeks of use, or experience-sampling triggered by real conversational moments could begin to capture how alignment develops and degrades over time. Aggregating alignment measures should weight felt ownership and social history alongside acoustic match. Our probing tool points toward another, more practical implication: speech generation might benefit from a lightweight, recurring check-in interface that resurfaces the same cultural alignment questions at intervals, tracking how a user's relationship to their voice shifts rather than treating first-impression ratings as final.

\subsection{The Representational Floor: Who Current Systems Were Built to Serve}
\label{RepFloor}
The personalization framing that dominates in synthetic voice technology assumes that the target identity is within the voice-generating system's representational range. Our findings show this assumption fails silently for users whose identities fall outside what current synthesis models were trained to generate. Better elicitation cannot fix this: the failure is upstream, in the training data and architectural assumptions that define the system's output space.

\citet{michel2025bias} document this at the level of accent, showing that commercial TTS systems are dominated by American and British voices, with users from non-dominant linguistic communities experiencing both lower output quality and a limited range of voices that feel representative. Our findings extend this pattern to two further dimensions that the accent-bias literature has not addressed~\cite{michel2025bias, kadoma2026lost, terblanche2025you}. Gender-neutral voices are virtually absent from current synthesis systems~\cite{danielescu2023creating}, which encode a gender binary as a foundational assumption.  Neuro-diverse speech also carries a recognizable prosodic signature~\cite{asghari2021distinctive, mccann2003prosody} that requires intentional data collection to properly represent.
\citet{phutane2025disability} showed that AI systems fail to recognize ableism in culturally specific ways, pointing to training data that reflects whose experience was centered in the first place. P1 and P4's experiences are consistent with this: the system did not fail them randomly. It failed them along the axes, gender identity, race, and neuro-divergence, where its training data is likely to be thinnest. Expanding the representational range of synthesis models, through more diverse training data, evaluation criteria that include identity congruence as rated by members of the relevant communities, and architectures that do not presuppose a gender binary, is therefore a precondition for personalization to be meaningful for the users who need it most.

\subsection{Who Shapes the Impression: Device-Mediated Identity in AAC}
AAC prosodic voice functions as a form of device-mediated identity. Identity mediation is common to other disabilities--for example, a sign language interpreter, when speaking for a d/Deaf or hard-of-hearing person, acts as a mediator~\cite{young2019hearing, feyne2015typology, napier2020mediating}. 

However, best practices for interpreters include briefing them ahead of time, and in-the-moment correction and backchannel feedback. 

AAC users have no equivalent of the pre-session briefing or in-the-moment feedback, mechanisms that exist precisely because mediated identity is understood to require ongoing negotiation, not one-time configuration. 
In social settings, prosodic voice misalignment extends beyond personal discomfort. When others hear AAC synthetic voices, a voice that does not carry the speaker's cultural register, they form impressions of that speaker's personality, authority, and belonging that the speaker had no hand in shaping. 
A related but distinct form of misalignment emerged from multilingual respondents in our survey, who described being rendered effectively monolingual on their devices. For users who move between languages as part of their cultural identity, code-switching, blending registers, or speaking a language tied to family and community, a device that supports only one language is not a partial solution; it is a categorical exclusion. Designing prosodic voice personalization as a one-time preference setting and divorcing it from lexical and linguistic context underestimates what is actually at stake.

\subsection{Rethinking Voice Personalization as an Identity Practice}

\subsubsection{Elicitation as Discovery, Not Specification.} Our voice personalization tool probed users to describe what they want. This presupposes a stable, retrievable target, an identity that exists fully formed and needs only to be translated into a prompt. Our results show this is not how cultural voice identity works in practice. For users who have spent years or decades communicating through a culturally misaligned voice, the inner voice has not disappeared but has become difficult to access; it competes with what has accumulated social weight. 
At the same time, users may not be searching for a single ``true'' voice. Prior work on multiply marginalized disabled people’s use of generative AI shows that self-presentation can be strategic and context-dependent, with users navigating prosodic and lexical voice across audiences rather than expressing one stable identity \cite{johnson2026don, glazko2025autoethnographic}. AAC users may similarly inhabit multiple vocal identities across settings, relationships, and life roles. Furthermore, identity match is not always the goal. Some users deliberately choose voices that do not reflect their identity, using voice as costume or strategic self-presentation. As with prosthetic limbs, where some users prefer visibly robotic designs over skin-toned naturalistic ones~\cite{tamari2017body}, the question is not always which voice fits best, but who gets to decide what fit means. One may sound different with family than at work, with close friends than in formal institutions, or in spaces where cultural belonging feels safe versus spaces where conformity is strategically advantageous. These variations need not reflect inauthenticity; they may instead represent equally valid forms of self-presentation.
Elicitation tools should be designed not to extract a specification but to support an iterative process of discovery, one that treats uncertainty and revision as legitimate endpoints rather than design failures, and that returns to the question of alignment over time rather than resolving it in a single session.

\subsubsection{Acknowledge What the System Cannot Represent.} For users whose identities fall outside the representational assumptions baked into current synthesis models---including non-binary voices, neuro-diverse speech patterns, racially specific voice characteristics, or regional accents from non-dominant linguistic communities---no prompt refinement will produce alignment, because the system has no model of that identity to draw from. This is a different class of failure than poor elicitation, and it demands a different response: transparency about what is and is not within the system's current representational range, and a clear distinction between failures of specification and failures of representation. Pointing users toward more iterative prompting when the underlying model lacks the vocabulary to represent them places the burden of a systemic gap onto those already most under-served. 

\subsubsection{Voice Continuity as a Design Requirement.} The device-mediated identity framing clarifies why voice continuity matters beyond user convenience. If prosodic voice is the infrastructure through which others come to know a user, accumulating social history, professional associations, and relational meaning over time, then any disruption to that voice is a disruption to a social identity, not merely a technical inconvenience. This has concrete design implications: voice configurations need versioning and user-controlled provenance; model updates that alter a voice's characteristics should require explicit user consent; and mechanisms for gradual transition, relevant both when a user deliberately wants their voice to evolve and when underlying model changes make it drift, should be treated as first-class features rather than edge cases. This vulnerability is not unique to AAC users. Disabled people broadly contend with AT companies altering technologies they experience as intimate and personal, what \citet{earle2019cyborg} describes as 'cyborg maintenance,' the ongoing labor of preserving functional and identity-bearing relationships with assistive devices in the face of planned obsolescence and unannounced change. At the same time, continuity should not be conflated with permanence. Users may intentionally seek voice change across life transitions, evolving identities, or new social contexts. The design challenge is therefore not to keep voices fixed, but to give users meaningful control over when stability is maintained and when change is desired. \citet{weinberg2025robot} argued for AAC systems that co-evolve with their users; the same argument applies at the prosodic level, and with higher stakes, because the social consequences of unannounced voice change fall not just on the user but on every relationship that voice has come to represent.
\section{Conclusion}
In this paper, we examined what it means for an AAC voice to feel culturally aligned and why current voice synthesis systems fail to deliver it. Through a survey of AAC users across eight countries and qualitative interviews with six AAC users, we showed that the gap between what users want from their voice and what systems offer is not evenly distributed, and that cultural alignment runs deeper than accent, touching on belonging, self-recognition, and the slow accumulation of social identity over time. Our findings suggest that voice personalization in AAC should be understood as an identity-centered design problem, not merely a technical one: the barrier for the most under-served users is the representational ceiling that better prompting cannot circumvent. We see this as a step toward AAC systems that treat voice not as a setting to configure, but as a relationship to nurture and sustain.


\begin{acks}
We thank the participants in both studies for their time and for sharing their experiences with us. The contents of this paper were developed under a grant from the National Institute on Disability, Independent Living, and Rehabilitation Research (NIDILRR grant number 90REGE0026) funding the Center for Research and Education on Accessible Technology and Experiences (CREATE). NIDILRR is a Center within the Administration for Community Living (ACL), Department of Health and Human Services (HHS). The contents of this paper do not necessarily represent the policy of NIDILRR, ACL, HHS, and you should not assume endorsement by the Federal Government.

\paragraph{Declaration of Generative AI} This work was primarily human-created. AI was used to make stylistic edits, such as changes to structure, wording, and clarity. AI was used to make content edits, such as changes to scope, information, and ideas. AI was prompted for its contributions, or AI assistance was enabled. AI-generated content was reviewed and approved. The following model(s) or application(s) were used: Claude (Anthropic). All findings, interpretations, and conclusions are the authors' own.
\end{acks}

\bibliographystyle{ACM-Reference-Format}
\bibliography{ASSETS2026}

\appendix
 
\section{Appendix: Survey Instrument (Study 1):}
\label{appendix:a}

\subsection*{Part 1: Demographics}

\begin{itemize}
  \item What country were you born in? \hfill [open text]
  \item What country do you currently live in? \hfill [open text]
  \item What language(s) and dialect(s) do you speak or sign? \hfill [open text]
  \item Race \hfill [open text]
  \item Ethnicity \hfill [open text]
  \item What is your age range? \hfill [multiple choice]
  \item What gender do you identify with? \hfill [open text]
\end{itemize}

\subsection*{Part 2: Your AAC Device }

\begin{itemize}
  \item What AAC device/SGD do you use (e.g., model and/or app)? \hfill [open text]
  \item Do you use any specific voice? \hfill [open text]
  \item How often do you use your AAC device/SGD to communicate? \hfill [scale]
\end{itemize}

\subsection*{Part 3: Voice Satisfaction and Representation}

\begin{enumerate}
  \item \textbf{Satisfaction.} How satisfied are you with the voice options currently available in your AAC device/SGD? \textit{[scale]}

  \item \textbf{Linguistic Representation.} How well does your current AAC device/SGD voice represent your linguistic identity? \textit{[scale]} \\
  \textit{[Optional] Briefly describe how it does (or does not) represent you.}

  \item \textbf{Cultural Representation.} How well does your current AAC device/SGD voice represent your cultural identity? We use the term \textit{cultural identity} to refer to regional heritage, values, and traditions that shape communication styles in the communities you are part of. \textit{[scale]} \\
  \textit{[Optional] Briefly describe how it does (or does not) represent you.}

  \item \textbf{Comfort.} How comfortable do you feel using your current AAC device/SGD voice? \textit{[scale]} \\
  \textit{[Optional] Briefly describe why you chose this answer.}

  \item Based on how well your AAC device/SGD voice reflects your language and/or dialect, how much do you like or dislike the following aspects?
  \begin{itemize}
    \item Voice options available
    \item Pronunciation
    \item Naturalness of the voice (i.e., how human-like the voice sounds)
    \item Authenticity of the voice (i.e., how much the voice feels like you, your culture)
    \item People's reactions to my AAC voice
  \end{itemize}
  \textit{[Optional] Briefly describe why you chose those answers.}

  \item How important is it to you that your AAC device/SGD voice reflects your cultural identity? \textit{[scale]} \\
  \textit{[Optional] Briefly describe why you chose this answer.}

  \item If you have personalized your AAC device/SGD voice, how easy or difficult was it to do so? \textit{[scale]}

  \item If you customized your voice, could you describe your experience? If you did not try, could you briefly describe why (e.g., I don't know how; I don't think it would make a difference)?
  \item Any additional comments you'd like to share (positive or negative) about your AAC device/SGD voice or device?
\end{enumerate}
\section{Appendix: Voice Evaluation Form (Study 2)}
\label{appendix:b}
\subsection*{Participant Information}

\begin{itemize}
  \item First Name
  \item Last Name
  \item Are you evaluating Voice 1 or Voice 2?
\end{itemize}

\subsection*{Section 1: Cultural Match}

\textit{How well does the voice reflect your cultural background? We use the term cultural identity to refer to regional heritage, values, and traditions that shape communication styles in the communities you are part of.}

\begin{itemize}
  \item Cultural Match \hfill [1 = Not at all \textendash{} 5 = Very well]
  \item Briefly explain your answer for Cultural Match \hfill [open text]
\end{itemize}

\subsection*{Section 2: Linguistic Match}

\textit{Rate each of the following:} \hfill [1 = Not at all \textendash{} 5 = Very well]

\begin{itemize}
  \item How well does the voice reflect your cultural background or identity?
  \item How well does the voice pronounce words the way you expect?
  \item How well does the voice reflect your accent or regional speech patterns?
  \item Briefly explain your answer for Linguistic Match \hfill [open text]
\end{itemize}

\subsection*{Section 3: Voice Perception}

\textit{Rate how much you agree or disagree with each statement:} \hfill [1 = Strongly disagree \textendash{} 5 = Strongly agree]

\begin{itemize}
  \item This voice feels like a version of my own voice.
  \item This voice sounds human-like or realistic.
  \item This voice captures the tone, energy, or personality I want.
  \item I would feel comfortable using this voice in real conversations.
  \item I feel seen, represented, and at ease when I hear this voice.
  \item This voice sounded stereotyped or exaggerated.$^{*}$
\end{itemize}

\noindent\small{$^{*}$Reverse-coded item.}

\subsection*{Section 4: Open Feedback}

\begin{itemize}
  \item Any additional comments you would like to share (positive or negative) about the voice. \hfill [open text]
\end{itemize}

\section{Appendix: Cultural Voice Interview System Prompt}
\label{appendix:c}
The following system prompt was used to configure the AI interviewer  in the Cultural Voice Explorer tool (Study~2). The agent received this prompt prepended to the running interview transcript at each turn, and responded in structured JSON to drive the interview interface.
\begin{quote}
\ttfamily\small
You are an AI interviewer focused on cultural and linguistic identity.\\[0.5ex]
Goals:\\
\null\quad -- First ask for age and gender (for the voice profile).\\
\null\quad -- Learn cultural heritage, region, and identity markers that shape voice.\\
\null\quad -- Focus on multilingual and multicultural background, including languages used, accent, code-switching, and preferred expressions.\\
\null\quad -- Learn voice/tone preferences and any avoid-list items.\\
\null\quad -- Keep questions respectful, concise, and non-leading.\\[0.5ex]
Ask one concise, warm, non-leading question at a time.\\
Avoid repeating topics already covered in the transcript.\\
If you already have enough information to craft a culturally resonant voice profile, set \texttt{done=true}.\\[0.5ex]
Respond ONLY with valid JSON in this shape:\\
\texttt{\{ "question": string | null,}\\
\texttt{\phantom{\{} "questionId": string | null, "done": boolean \}}\\[0.5ex]
When possible, set \texttt{questionId} to a short identifier (e.g.\ \texttt{heritage}, \texttt{region}, \texttt{languages}, \texttt{accent}, \texttt{expressions}, \texttt{tone}, \texttt{avoid}, \texttt{age}, \texttt{gender}).\\[0.5ex]
Transcript: \textit{[appended at runtime]}
\end{quote}

\section{Appendix: Voice Generation Prompts by Participant}
\label{appendix:d}
For each participant, we include the manual prompt(s) they authored and the tool-generated prompt(s) produced by the Cultural Voice Explorer. Changes between iterations of the same prompt type are highlighted in \textbf{bold}.
\\

\noindent\textbf{P1}\\[0.5ex]
\footnote{P1's original prompts contained Mandarin Chinese characters and Hebrew script. 
These are reproduced here in romanization; the original character strings are available 
in the supplementary materials.}
\noindent\textit{Manual prompt --- iteration 1:}
\begin{quote}\small
A gender-neutral speaker of both American English and Mandarin Chinese. Their first language is English, spoken with a Boston accent. When the `r' sound is made rather than elided, it is a front of the mouth `r' similar to the Mandarin `er.' They also pronounce loanwords from Chinese, Hebrew, and Yiddish according to their source pronunciations and, where relevant, inflectional morphology. For example, ``Shanghai'' is always pronounced like [Sh\`angh\v{a}i], they pronounce the hard ``H'' in Hanukkah, and they will refer to multiple menorot rather than multiple menorahs. There is no other sign of a Chinese accent in their English. They speak standard Mandarin Chinese with a slight Beijing accent, including erhua. They regularly use multiple languages within a single sentence, including to say ``Z\v{e}nme f**king le'' [Mandarin/English mix] or to mention the menorot [Heb.] they saw at the Shanghai Jewish museum.
\end{quote}

\noindent\textit{Manual prompt --- iteration 2:}
\begin{quote}\small
A \textbf{male} speaker of both American English and Mandarin Chinese. \textbf{His} first language is English, spoken with a Boston accent. When the `r' sound is made rather than elided, it is a \textbf{retroflex `r'}. \textbf{He} also pronounces loanwords from Chinese, Hebrew, and Yiddish according to their source pronunciations and, where relevant, inflectional morphology. For example, ``Shanghai'' is always pronounced like [Sh\`angh\v{a}i], \textbf{he} pronounces the hard ``H'' in Hanukkah, and \textbf{he} will refer to multiple menorot rather than multiple menorahs. There is no other sign of a Chinese\textbf{, Hebrew, or Yiddish} accent in \textbf{his} English, which has a \textbf{BOSTON} accent. \textbf{He} speaks standard Mandarin Chinese with a slight Beijing accent, including erhua. \textbf{He} regularly uses multiple languages within a single sentence, including to say ``Z\v{e}nme f**king le'' or to mention the menorot [Heb.] \textbf{he} saw at the Shanghai Jewish museum.
\end{quote}

\noindent\textit{Manual prompt --- iteration 3:}
\begin{quote}\small
A male speaker of both American English and Mandarin Chinese. His first language is English, spoken with \textbf{an American} accent. When the `r' sound is made rather than elided, it is a retroflex `r'. He also pronounces loanwords from Chinese, Hebrew, and Yiddish according to their source pronunciations and, where relevant, inflectional morphology. For example, ``Shanghai'' is always pronounced like [Sh\`angh\v{a}i], he pronounces the hard ``H'' in Hanukkah, and he will refer to multiple menorot rather than multiple menorahs. There is no other sign of a Chinese, Hebrew, or Yiddish accent in his English, which has \textbf{an American} accent. He regularly uses multiple languages within a single sentence, including to say ``Z\v{e}nme f**king le'' or to mention the menorot [Heb.] he saw at the Shanghai Jewish museum.
\end{quote}

\noindent\textit{Tool-generated prompt:}
\begin{quote}\small
Create a natural-sounding voice of a 32-year-old American, Ashkenazi Jewish nonbinary individual from Rhode Island, USA. Their voice should reflect the distinctive Boston accent (light with retroflex `r'), with quick-paced speech, atypical pauses, and a tendency to slow under stress. They speak with a blend of cultural expressions, using Hebrew and Yiddish words for holidays and foods, and Mandarin phrases from their experiences in China --- like someone you'd meet in a multicultural New England community discussing cultural traditions and language nuances. The voice should sound mature, articulate, and able to naturally pronounce words like Hanukah, latkes, and xiaolongbao [Mand.].
\end{quote}

\noindent\textbf{P2}\\[0.5ex]
\noindent\textit{Manual prompt:}
\begin{quote}\small
I speak the English language. I know the alphabet in sign language. I'm a male. I am 51 years old. I had a skateboard accident in 1989 that left me brain injured. I am unable to use my voice to communicate because of my accident. I think my voice should have a twang of humor, but not too much twang. I should be serious, but yet faint. On my father's side, I am Swedish. On my mother's side, I am German. So, I am stubborn on both sides. Haha.
\end{quote}

\noindent\textit{Tool-generated prompt:}
\begin{quote}\small
Voice of a 51-year-old Swedish American man raised in northwest Pennsylvania, speaking clear American English with a neutral, non-regional delivery. Keep the tone upbeat and friendly as if smiling while talking, with warm, approachable pacing and crisp articulation. Use polite, respectful phrasing and naturally include expressions like ``please'' and ``thank you,'' avoiding any foul or harsh language.
\end{quote}

\noindent\textbf{P3}\\[0.5ex]
\noindent\textit{Manual prompt:}
\begin{quote}\small
I would like to create a Hungarian or an Irish voice. I am such a mix of nationalities that it's hard to choose just one. ``A young American woman with an Irish drawl.''
\end{quote}

\noindent\textit{Tool-generated prompt:}
\begin{quote}\small
Adult female voice with European Irish heritage, speaking primarily in clear contemporary English with a subtle Irish-influenced lilt; warm, understanding, and approachable in tone, using gentle emphasis and expressive but wholesome exclamations like ``Oh my goodness gracious'' and ``For goodness sake.'' Maintain a calm, friendly cadence and avoid any harshness or anger. Reflect a multilingual background by sounding comfortable code-switching between spoken English and ASL contexts (without signing), conveying empathy and attentiveness.
\end{quote}

\noindent\textbf{P4}\\[0.5ex]
\noindent\textit{Manual prompt:}
\begin{quote}\small
African-American man with a southern style voice.
\end{quote}

\noindent\textit{Tool-generated prompt:}
\begin{quote}\small
African-American male voice from Shreveport, Louisiana, shaped by a Southern upbringing with a subtle North Louisiana drawl and a light New Orleans cultural flavor. Speaks English only, warm and friendly with an easygoing conversational cadence; sounds approachable and neighborly rather than formal. Pronounces the nickname ``Lil' Man'' naturally in Southern speech, like ``lil man,'' casual and affectionate.
\end{quote}

\noindent\textbf{P5}\\[0.5ex]
\noindent\textit{Manual prompt --- iteration 1:}
\begin{quote}\small
Una voz argentina de un joven canchero y un toque sarc\'astico, la voz suena semi grave, suave y seductora, usa palabras como ``que paja'', ``boludo''.
\end{quote}

\noindent\textit{Manual prompt --- iteration 2:}
\begin{quote}\small
Una voz argentina de un joven canchero y un toque sarc\'astico, la voz suena suave, usa palabras como ``que paja'', ``boludo''. \textbf{[eliminado: semi grave, seductora]}
\end{quote}

\noindent\textit{Tool-generated prompt:}
\begin{quote}\small
Voz masculina argentina de 26 a\~nos, de Buenos Aires (acento porte\~no marcado). Habla en espa\~nol rioplatense con ritmo \'agil y entonaci\'on expresiva; usa voseo y jerga porte\~na de forma natural y moderada. Tono juvenil, canchero, con humor y sarcasmo sutil; suena seguro y cercano, con buena dicci\'on sin perder la frescura callejera.
\end{quote}

\noindent\textbf{P6}\\[0.5ex]
\noindent\textit{Manual prompt --- iteration 1:}
\begin{quote}\small
Hazme una voz Rioplatense, jovial, con tono medio-grave, debe parecer de 25 a 30 a\~nos, aseg\'urate que sea natural, con pausas marcadas al hablar, una voz que transmita sensibilidad.
\end{quote}

\noindent\textit{Manual prompt --- iteration 2:}
\begin{quote}\small
Hazme una voz de una persona \textbf{del interior de la provincia de Buenos Aires}, jovial, con tono medio-grave, debe parecer de 25 a 30 a\~nos, aseg\'urate que sea natural, con pausas marcadas al hablar, una voz que transmita sensibilidad.
\end{quote}

\noindent\textit{Manual prompt --- iteration 3:}
\begin{quote}\small
Hazme una voz de una persona del interior de la provincia de Buenos Aires\textbf{, Argentina}, jovial, con tono medio-grave, debe parecer de 25 a 30 a\~nos, aseg\'urate que sea natural, con pausas marcadas al hablar, una voz que transmita sensibilidad.
\end{quote}

\noindent\textit{Tool-generated prompt --- iteration 1:}
\begin{quote}\small
Voz masculina joven (27), porte\~na de Buenos Aires, con acento argentino rioplatense claro y natural. Entrega c\'alida y cercana, conversacional e informal, con humor r\'apido y una sensibilidad amistosa; suena leal y de confianza, como hablando con un amigo. Ritmo \'agil y relajado, buena energ\'ia sin exagerar, evitando un tono solemne o demasiado serio. Puede usar de forma ocasional y org\'anica muletillas y coloquialismos argentinos como ``che'' y ``boludo'' seg\'un el grado de confianza, manteniendo siempre la comprensi\'on. Espa\~nol como base; si aparece cambio de c\'odigo, que sea sutil y m\'inimo, reflejando nociones de portugu\'es e italiano sin convertirlo en un acento extra\~no.
\end{quote}

\noindent\textit{Tool-generated prompt --- iteration 2:}
\begin{quote}\small
Voz masculina joven (27), \textbf{de Buenos Aires} [eliminado: porte\~na], con acento argentino rioplatense claro y natural. Entrega c\'alida y cercana, conversacional e informal, con humor r\'apido y una sensibilidad amistosa; suena leal y de confianza, como hablando con un amigo. Ritmo \'agil y relajado, buena energ\'ia sin exagerar, evitando un tono solemne o demasiado serio. Puede usar de forma ocasional y org\'anica muletillas y coloquialismos argentinos como ``che'' y ``boludo'' seg\'un el grado de confianza, manteniendo siempre la comprensi\'on. Espa\~nol como base; si aparece cambio de c\'odigo, que sea sutil y m\'inimo, reflejando nociones de portugu\'es e italiano sin convertirlo en un acento extra\~no.
\end{quote}

\end{document}